\documentclass[acmsmall]{acmart}
\usepackage{makecell}
\usepackage{ragged2e}
\usepackage{float}
\usepackage{xurl}
\usepackage{hyperref}
\usepackage{breakurl}

\AtBeginDocument{%
  \providecommand\BibTeX{{%
    \normalfont B\kern-0.5em{\scshape i\kern-0.25em b}\kern-0.8em\TeX}}}
    
\makeatletter
\renewcommand\@makefntext[1]{\leftskip=1em\hskip-2em\@makefnmark#1}
\makeatother

\setcopyright{rightsretained}
\acmJournal{PACMHCI}
\acmYear{2021} \acmVolume{5} \acmNumber{CSCW2} \acmArticle{433} \acmMonth{10} \acmPrice{}\acmDOI{10.1145/3479577}

\received{January 2021} 
\received[revised]{April 2021}
\received[revised]{July 2021}
\received[accepted]{July 2021}

\begin{document}
\newcommand{\change}[1]{{\textcolor{blue}{#1}}}

\title[Everyday Algorithm Auditing]{Everyday Algorithm Auditing: Understanding the Power of Everyday Users in Surfacing Harmful Algorithmic Behaviors}


\author{Hong Shen}
\authornote{Co-first authors contributed equally to this research.}
\email{hongs@andrew.cmu.edu}
\author{Alicia DeVos}
\authornotemark[1]
\email{adevos@andrew.cmu.edu}
\affiliation{
\institution{Carnegie Mellon University}
\streetaddress{5000 Forbes Ave}
\city{Pittsburgh}
\state{PA}
\country{USA}
\postcode{15213}
}

\author{Motahhare Eslami}
\authornote{Co-senior authors contributed equally to this research.}
\email{meslami@andrew.cmu.edu}
\author{Kenneth Holstein}
\authornotemark[2]
\email{kjholste@andrew.cmu.edu}
\affiliation{%
\institution{Carnegie Mellon University}
\streetaddress{5000 Forbes Ave}
\city{Pittsburgh}
\state{PA}
\country{USA}
\postcode{15213}
}


\begin{abstract}
A growing body of literature has proposed formal approaches to audit algorithmic systems for biased and harmful behaviors. While formal auditing approaches have been greatly impactful, they often suffer major blindspots, with critical issues surfacing only in the context of everyday use once systems are deployed. Recent years have seen many cases in which \emph{everyday users} of algorithmic systems detect and raise awareness about harmful behaviors that they encounter in the course of their everyday interactions with these systems. However, to date little academic attention has been granted to these bottom-up, user-driven auditing processes. In this paper, we propose and explore the concept of \textit{everyday algorithm auditing}, a process in which users detect, understand, and interrogate problematic machine behaviors via their day-to-day interactions with algorithmic systems. We argue that everyday users are powerful in surfacing problematic machine behaviors that may elude detection via more centrally-organized forms of auditing, regardless of users’ knowledge about the underlying algorithms. We analyze several real-world cases of everyday algorithm auditing, drawing lessons from these cases for the design of future platforms and tools that facilitate such auditing behaviors. Finally, we discuss work that lies ahead, toward bridging the gaps between formal auditing approaches and the organic auditing behaviors that emerge in everyday use of algorithmic systems. 
\end{abstract}



\maketitle
\pagestyle{plain}

\section{Introduction}
Algorithmic systems increasingly exercise power over many aspects of our everyday lives, including which advertisements or social media posts we see, the quality of healthcare we receive, the ways we are represented to our peers or potential employers, and which neighborhoods are subjected to increased policing \cite{seaver2019knowing,gillespie2014relevance,diakopoulos2014algorithmic,noble2018algorithms, veale2018fairness}. These systems, however, are not infallible. A growing body of research has drawn attention to the societal impacts of algorithmic systems’ behaviors, examining the ways these systems can, whether inadvertently or intentionally, serve to amplify existing biases and social inequities or create new ones (e.g., \cite{buolamwini2018gender,noble2018algorithms,sweeney2013discrimination,zou2018ai}). In response to these concerns, scholars have proposed a number of approaches to audit algorithmic systems for biased, discriminatory, or otherwise harmful behaviors\footnote{Throughout this paper, we distinguish between harmful algorithmic biases and harmful algorithmic behaviors more broadly, as appropriate. Although these categories often overlap, we acknowledge that not all algorithmic harms are best understood as ``biases,’’ and not all algorithmic biases are necessarily harmful \cite{blodgett2020language}.}\thinspace—\thinspace both internally, within the organizations responsible for developing and maintaining these systems, and externally \cite{sandvig2014auditing,raji2020closing}. 

Algorithm auditing approaches that have been discussed in the literature often occur outside the context of everyday use of an algorithmic system or platform: auditors, such as researchers, algorithmic experts, and activists, initiate, conduct, or orchestrate the auditing process. While such audits have made significant contributions in detecting biases and harmful algorithmic behaviors, they often fail to surface critical issues. For example, even after devoting months of effort to internal auditing, machine learning development teams often struggle to detect and mitigate harmful biases in their systems due to their own cultural blindspots \cite{holstein2019improving}. Additional reasons why formal auditing approaches can fail to detect serious issues include the presence of unanticipated circumstances or social dynamics in the contexts where a system is used, as well as changing norms and practices around the use of algorithmic systems over time \cite{friedman1996bias}. These factors make detecting certain harmful algorithmic behaviors difficult for auditors unless they are embedded within particular real-world social or cultural contexts \cite{cramer2018assessing,holstein2020replay,seaver2017algorithms,young2019toward}. 

While existing formal audits often struggle to detect critical issues in algorithmic systems, recent years have seen many cases in which \emph{users} of algorithmic systems detect and raise awareness about biased and harmful behaviors that they encounter in the course of their everyday interactions with algorithmic systems. 
One recent example is the highly publicized case of Twitter’s image cropping algorithm exhibiting racial discrimination by focusing on white faces and cropping out Black ones. Twitter users began to spot issues around this algorithm and came together organically to investigate. Through online discussions, they built upon one another’s findings to surface similar biases or to present evidence or counter-evidence for a pattern discovered by another person. This occurred even though the company stated that “it had tested the service for bias before it started using it” \cite{guardian2020}. Twitter’s testing procedures failed to detect this bias because during real world usage, users were interacting with the cropping algorithm in ways the team did not anticipate up front.

Although similar auditing behaviors have been observed around a range of algorithmic systems, including image search engines, machine translation, online rating/review systems, image captioning, and personalized advertising, many open questions remain regarding how users of algorithmic systems come to detect, report, and theorize about harmful algorithmic biases and behaviors in their day-to-day use of a platform. What factors facilitate “successful” user-driven algorithmic audits, and what factors prevent such audits from having an impact? How might we effectively harness everyday users’ motivation and strengths in surfacing harmful algorithmic behaviors, so as to overcome the limitations of existing auditing approaches?

In this paper, we take a first step toward answering these questions by proposing and exploring the concept of \emph{everyday algorithm auditing:} a process in which users detect, interrogate, and understand problematic machine behaviors via their daily interactions with algorithmic systems. In contrast to formal auditing approaches, everyday algorithm auditing occurs in the context of ``everyday use’’ of an algorithmic system or platform. Accordingly, throughout this paper, we use the term ``everyday users’’ to refer to social actors who take part in everyday use of an algorithmic system. We draw insights from past literature in “everyday resistance” \cite{de1984practice,scott1985weapons} and ``algorithmic resistance’’ \cite{velkova2019algorithmic} to theorize such behaviors. We interpret everyday algorithmic auditing efforts as a form of everyday resistance by which users actively, continuously question and repurpose mass cultural products in their everyday lives to resist the hegemonic culture forms of their times. In resisting algorithmic harms and conducting everyday audits, everyday users usually form counterpublics, ``parallel discursive arenas’’ \cite{fraser1990rethinking}, where communities impacted by harmful algorithmic behaviors come together and participate in their own forms of collective sensemaking, hypothesis development, and algorithm bias detection. 

We adopt an exploratory case study approach, examining several recent cases to understand the nature and dynamics of everyday algorithm audits. In particular, we examine two types of algorithmic sociotechnical platforms, the Twitter cropping algorithm and online rating algorithms, that have been the target of everyday algorithm audits in recent years. First, we analyze the Twitter racial bias case mentioned above \cite{guardian2020}. Second, we look at another everyday auditing process investigating potential gender bias caused by the same algorithm.\footnote{\url{https://twitter.com/mraginsky/status/1080568937656565761}} Third, we turn our attention to a case surrounding Yelp’s review filtering algorithm, in which a large group of small business owners came together to verify its bias against businesses that do not advertise with Yelp \cite{Kang2013}. Finally, we examine a case surrounding Booking.com’s online rating algorithm, wherein users came together to audit the system after noticing that the ratings calculated by the algorithm did not match their expectations \cite{eslami2017careful}. In all of these cases, we compare the different paths everyday users took to detect, examine, discuss, and understand biases in algorithmic systems’ behavior.

Analyzing these cases and comparing their dynamics led us to a process-oriented view of everyday algorithm audits encompassing (1) initiation of an audit to (2) raising awareness of observed issues to (3) hypothesizing about observed behaviors and testing an algorithmic system to ideally (4) some form of remediation. Although some everyday audits may terminate before touching all of these phases, this process-oriented view provides a useful high-level understanding for describing the paths everyday audits can take, comparing different everyday audits, and envisioning new kinds of interventions to facilitate such audits. We argue that day-to-day users are powerful in surfacing problematic machine behaviors that may elude detection via existing, formal auditing approaches, even when users lack technical knowledge of the underlying algorithms. Furthermore, we suggest that everyday algorithm auditing may be most powerful when audits are conducted collectively, through interactive discussions among users with a diverse range of experiences.

Our case analysis and process model of everyday algorithm audits inform a series of design implications for future work toward supporting more effective everyday algorithm auditing. We outline five broad categories of potential design interventions to support everyday audits\thinspace—\thinspace (a) community guidance, (b) expert guidance, (c) algorithmic guidance, (d) organizational guidance, and (e) incentivization\thinspace—\thinspace and discuss potential design trade-offs. Finally, we close with a discussion of work that lies ahead, in order to bridge the gap between existing algorithm auditing approaches in academia and industry versus everyday auditing behaviors that emerge in day-to-day use of algorithmic systems.

\section{Algorithm Auditing}
In this section, we briefly overview existing approaches to algorithm auditing, an area that has received increasing attention from the CSCW, HCI, and ML communities in recent years, and describe how our work contributes to this emerging line of research.

\subsection{Existing Approaches to Algorithm Auditing}
Scholars have proposed a number of approaches to audit algorithmic systems for biased, discriminatory, or otherwise harmful behaviors, often under the broad umbrella of “auditing algorithms” \cite{sandvig2014auditing}. Drawing from a wide range of methods in areas such as investigative journalism, information security and finance, and housing and employment audits, this line of work often involves third-party external experts (e.g., researchers, technologists, or policymakers) in assessing the alignment of deployed algorithmic systems with laws and regulations, societal values, ethical desiderata, or industry standards (e.g., \cite{buolamwini2018gender,diakopoulos2014algorithmic,raji2019actionable,sandvig2014auditing}). Meanwhile, a growing body of work has proposed tools, processes, and frameworks for internal algorithm audits, conducted by ML teams themselves, with the aim of detecting and mitigating misalignments prior to deployment (e.g., \cite{cramer2018assessing,holstein2019improving,lee2020landscape,raji2020closing}).

Past research in this domain has uncovered harmful biases across a wide range of algorithmic systems such as search engines \cite{robertson2018auditing, noble2018algorithms,kulshrestha2017quantifying}, housing websites \cite{asplund2020auditing}, hiring systems \cite{dastin2018amazon}, online employment services \cite{chen2018investigating} and e-commerce platforms \cite{hannak2014measuring}. For example, Robertson et al. used a survey and web browser extension to collect users’ experiences in order to audit partisan audience biases in Google search results \cite{robertson2018auditing}. In another example, Hannak et al. used web scraping techniques and hired Amazon MTurk users as testers in order to audit a variety of top e-commerce sites for price steering and discrimination \cite{hannak2014measuring}. These are just a few examples of many auditing efforts that have been conducted during recent years on algorithmic systems. 

Inspired by early methods of the social scientific “audit study” \cite{mincy1993urban}, Sandvig et al. \cite{sandvig2014auditing} proposed a taxonomy to summarize different algorithm auditing methods and research designs, including (1) code audits, (2) noninvasive user audits, (3) scraping audits, (4) sock puppet audits, and (5) crowdsourced/collaborative audits. A \emph{code audit} typically requires auditors to have secure access to the source code and system design. A \emph{noninvasive user audit} might use methods like surveys to collect and synthesize users' experiences in order to support inferences about the underlying operations of an algorithmic system. A \emph{scraping audit} involves researchers sending out repeated queries to test how an algorithmic system behaves under a variety of conditions. In a \emph{sock puppet audit}, researchers generate fake user accounts, or “sock puppets,” to investigate how an algorithmic system may behave differently in response to different user characteristics or patterns of behavior. Finally, in a \emph{crowdsourced/collaborative audit}, instead of using “sock puppet” accounts, researchers might hire crowdworkers as testers via crowdsourcing platforms like Amazon MTurk to work on decomposed subtasks.

\subsection{Limitations of Existing Auditing Approaches}
Conducted as research projects or professional services, most algorithm auditing approaches that have been discussed in the literature require individuals with some level of technical expertise to initiate, conduct, or direct the entire process. For example, even approaches that rely on crowdsourcing techniques involve some centralized organization. In such crowdsourced/collaborative audits, it is still the researchers’ responsibility to design and initiate the audit by assigning crowdworkers specific tasks, and then to convert the crowd’s outputs on distributed tasks into meaningful insights about an algorithmic system’s behavior. However, such centrally-organized, formal audits often fail to surface serious issues that everyday users of algorithmic systems are quickly able to detect once a system is deployed in the wild. For example, Holstein et al. \cite{holstein2019improving} found that even when ML product teams thoroughly audit their systems using existing approaches, they often are subsequently blindsided by user complaints about myriad issues that their investigations had not revealed. In one case, even after devoting many months to identifying and mitigating gender and racial biases in their image captioning system, a product team learned about a wide array of additional issues that their efforts had ignored (e.g., mosques, synagogues, and other religious sites being labelled as “churches”) only after the system was deployed and in use across a range of real-world contexts.

Many harmful algorithmic behaviors are challenging to anticipate or detect outside of authentic, situated contexts of use, for a variety of reasons. For example, certain algorithmic behaviors may only arise\thinspace—\thinspace or may only be recognized as harmful\thinspace—\thinspace when a system is used in the presence of particular real-world social or cultural dynamics \cite{holstein2019improving,madaio2020co,seaver2017algorithms,selbst2019fairness}. However, these real-world dynamics may be difficult or infeasible to predict and simulate in an artificial setting. Other harmful behaviors may only emerge when a system is used in unanticipated ways or  in unanticipated contexts, perhaps due to changing norms and practices surrounding the use of a given algorithmic system over time \cite{cramer2018assessing,selbst2019fairness}. One highly publicized example is the Microsoft AI chatbot, Tay, which was set up to learn over time based on its interactions with users on Twitter. In less than 24 hours of exposure to Twitter users, who began interacting with Tay in ways its developers had not foreseen, the bot began exhibiting misogynistic and racist behaviors and ultimately needed to be shut down \cite{vincent2016chatbot}. These examples align with the broad notion of “emergent bias,” discussed in Friedman and Nissenbaum’s seminal work on bias in computer systems \cite{friedman1996bias}. 

Another reason why existing auditing approaches can fail to detect serious issues is that those involved in an audit may lack the relevant cultural backgrounds and lived experiences to recognize or know where to look for harmful behaviors \cite{young2019toward}. Although crowdsourced/collaborative auditing approaches invite participation from a larger number of people, such audits typically rely on crowdworkers, who do not necessarily represent the demographics of a given algorithmic system’s user base, and who are participating in structured tasks outside of their regular interactions with technology. Without engaging diverse, contextually situated users in the process of auditing complex algorithmic systems, existing approaches are likely to suffer major blindspots, with critical issues surfacing only post-deployment (e.g., \cite{cramer2018assessing,eslami2019user,friedman1996bias,holstein2019improving,selbst2019fairness}).

Despite the limitations of existing auditing approaches and the emergence of many everyday algorithm auditing efforts in recent years, this phenomenon remains understudied and has not yet been situated in the academic literature on algorithm auditing. In this paper, we investigate and characterize the concept of everyday algorithm auditing and describe how it compares with existing, formal algorithm auditing approaches. In particular, we emphasize the \textit{situatedness} \cite{suchman1987plans} of everyday algorithm audits within everyday use as a central factor that distinguishes this concept from other approaches.
In addition, we foreground that compared with existing crowd-based approaches (e.g., ``collaborative audits'' \cite{sandvig2014auditing}), which tend to be organized and managed centrally by a crowd-external party, everyday audits are more collectively organized and organic, and require less algorithmic expertise on the part of those organizing and enacting the audit.

Here we make a distinction between “collective” and the notion of “crowd” that is frequently used in the crowdsourcing literature. In a conventional “crowdsourced audit” \cite{sandvig2014auditing}, users are hired via crowdsourcing platforms to independently work on decomposed subtasks determined entirely by outside parties, often without clear understanding and control of the direction of the auditing process. In an everyday algorithm audit, however, users play a major role in deciding their own course of action, often collectively. Users’ autonomy and agency are central to the concept of everyday algorithm auditing.

\section{Everyday Algorithm Auditing}

\begin{table*}[t]
    \renewcommand{\captionfont}{\small}
    \renewcommand{\captionlabelfont}{\small}
   \fontsize{7}{7}\selectfont
    \setlength{\abovecaptionskip}{1pt}
    \renewcommand{\arraystretch}{1.1}
    \begin{tabular}{p{2 cm} p{3.3 cm} p{7.5 cm}}
      	\toprule
        \textbf{Domains} &\textbf{Cases} & \textbf{Descriptions}\\
        \midrule
    Search &  Google Image Search \cite{noble2018algorithms}  & 
Researcher Noble searched “black girls” on Google and found out the results were primarily associated with pornography.\\
    \midrule
    Rating/review  &  \textbf{Yelp advertising bias \cite{eslami2019user} }  & Many small business owners on Yelp came together to investigate Yelp’s potential bias against businesses that do not advertise with Yelp. \\
\\
  & \textbf{Booking.com quality bias} \cite{eslami2017careful}   &
A group of users on Booking.com scrutinized its rating algorithm after realizing the ratings appeared mismatched with their expectations.  \\
    \midrule    
Image cropping & \textbf{Twitter racial cropping} \cite{bbc2020twitter} & Researcher Madland posted an image on Twitter of himself and a Black colleague who had been erased from a Zoom call after Zoom’s algorithm failed to recognise his face.  Twitter automatically cropped the image to only show Madland. After Madland posted this too on Twitter, users joined him in testing the Twitter cropping algorithm.\\
\\
&  \textbf{Twitter gender cropping}$^{1}$  &A researcher noticed that when VentureBeat shared an article with an image of two women and two men on Twitter, the algorithm cropped out both women’s heads. This was followed by other reports and testing by Twitter users. \\
 \midrule
 Image captioning & ImageNet Roulette \cite{crawford2019excavating} & Artists/researchers Crawford and Paglen built an interface to help users test a model trained on the ImageNet dataset. Although technical experts provided a tool that supported everyday auditing behaviors, users had autonomy in directing their efforts and using images of their choosing. Collective auditing behaviors emerged through discussions on social media.\\
 \midrule
Image recognition & 
Google Photos \cite{guynn2015google} & Programmer Alciné uploaded photos to Google Photos and noticed the app labelled his Black friends as ``gorillas.''\\
 \midrule
 Advertising & Google’s online ad delivery \cite{sweeney2013discrimination} & Researcher Sweeney performed her classic auditing of online ad delivery after she found her own name presented as linked with an arrest record in one of her many typical searches on Google.\\
\midrule
Recommendation systems & YouTube LGBTQ+ demonetization \cite{romano2019group} & A group of YouTubers found that the YouTube recommendation algorithm demonetizes LGBTQ+ content, resulting in a huge loss of advertising revenue for LGBTQ+ content creators.\\
\\
 & Google Maps \cite{Fung2015map} & 
A group of users reported that when they searched for the N-word on Google Maps, it directed them to the Capitol building, the White House, and Howard University, a historically Black institution. Other users joined the effort and uncovered other errors.\\
 \\
 &  TikTok recommendation algorithm  \cite{KARIZAT2021TikTok,simpson2021you} & A group of users found that TikTok’s "For You Page" algorithm suppresses content created by people of certain social identities, including LGBTQ+ users and people of color. As a result, they worked together to amplify the suppressed content. \\
\midrule
Translation & Google Translate Quality-of-service harm$^{2}$ & 
Computer engineer Rezanasab noticed that Google Translate mistranslated ``condolences'' in Persian to ``congratulations'' in English when the phrase was directed toward people of certain countries, like Lebanon. In response, other users collectively tested the system using their own examples, showing this bias for some other nations, and discussed their observations.\\
\\
& Google Translate gender bias \cite{olson2018algorithm} & 
Social media users tested Google Translate for gender bias after noticing it associates certain genders with professions and activities when translating from gender-neutral languages such as Turkish, Hungarian, and Farsi. \\
\midrule 
Credit Card & Apple Card \cite{vigdor2019apple} & Tech entrepreneur Hansson noticed the credit limit of his Apple Card was 20 times higher than his wife’s, even though she has a higher credit score and they file joint tax returns.\\
\midrule
Facial recognition & Gender Shades \cite{buolamwini2018gender,buolamwini2016m} & Researcher Buolamwini noticed a problem when she was working with a facial analysis software: the software didn't detect her own face. She went on to evaluate the accuracy of a number of AI-powered gender classification products on people of different genders and skin tones.\\
    \bottomrule
    \end{tabular}
    \\
\flushleft{$^{1}$\url{https://twitter.com/mraginsky/status/1080568937656565761}. To avoid identifying the user who initially tweeted, as they have since \\\thinspace\thinspace\thinspace deleted their Twitter account, we avoid detailing this aspect of this case in depth.}\\

      
\flushleft{$^{2}$\url{https://twitter.com/Arnasab/status/1290956206232698880}.}
\\
 \caption{15 cases of everyday algorithm auditing across various domains, which we use to ground our discussion throughout this paper. Bold cases are analyzed in depth.}
     \label{tab:cases}
\end{table*}

We develop the concept of \emph{everyday algorithm auditing} as a general framework to capture and theorize a diverse set of emerging user behaviors around algorithmic systems. We define everyday algorithm auditing as the ways everyday users detect, understand, and/or interrogate problematic machine behaviors via their day-to-day interactions with algorithmic systems. Following \cite{sandvig2014auditing}, we use “auditing” to refer to behaviors undertaken, whether formally or informally, to test and investigate whether a given algorithmic system operates in a socially harmful way, such as behaving in ways that produce inequitable outcomes along lines of class, race, or gender. Building on Suchman \cite{suchman1987plans}, we argue that everyday algorithm auditing practices are situated actions in which everyday users encounter problematic machine behaviors via their routine interactions with algorithmic systems, then adjust their behaviors to understand and act on what they encountered.

Recent years have seen several cases of everyday algorithm auditing, with some publicized widely. However, apart from some notable exceptions \cite{eslami2019user,eslami2017careful}, to date little academic attention has been granted to characterizing these bottom-up, user-driven auditing behaviors. In this paper, we aim to bridge the gap between algorithmic auditing approaches proposed in the academic research literature and auditing behaviors that users exhibit day-to-day. In doing so, we explore what lessons can be drawn from the study of everyday algorithm auditing to overcome limitations of existing auditing approaches in the literature.

In this section, we first describe the scope and boundaries of everyday algorithm auditing. We then elaborate on how we develop the concept based on two streams of past literature, namely, everyday algorithmic resistance and counterpublics. As discussed below, we argue that everyday algorithm auditing can be viewed as a form of \emph{everyday resistance}. When performed by a group of users, everyday algorithm auditing can additionally be understood through the lens of \emph{counterpublics}, where members of often disadvantaged and marginalized social groups participate in their own form of collective sensemaking, opinion formation, and consensus building.

\subsection{Scope and Boundaries}
As an emerging phenomenon, the conceptual boundaries of everyday algorithm auditing remain somewhat fluid. In this paper, we scope this concept broadly, to include any actions taken by everyday users to notice, interpret, question, or bring attention to problematic algorithmic behaviors. Below we highlight some key aspects of the concept to help elucidate its scope: (1) what algorithmic expertise the everyday users have, (2) the extent to which an everyday audit represents a collective effort, and (3) how organic the entire process is. These three dimensions, reviewed below, are not intended to be exhaustive; rather, they serve as a starting point to better illuminate the definition. 

\textbf{Algorithmic Expertise:} While everyday algorithm auditing is defined by everyday use, this does not imply that the users engaging in these behaviors necessarily lack technological skills and algorithmic expertise. Sometimes, everyday audits are conducted by users who have a high degree of relevant technical knowledge. For example, Sweeney, who is a professor of the practice of government and technology, performed her classic audit of online ad delivery only after she found that her own name was linked with an arrest record during a casual search on Google \cite{sweeney2013discrimination}. In this case, Sweeney acted as a situated user while also having relevant technical expertise in the area. In another example, Noble, a professor of information studies and expert in algorithmic bias and discrimination, began her examination of racist bias in search engines after trying to find an engaging activity for her stepdaughter and nieces by googling “black girls.” Instead, her web search yielded pornography \cite{noble2018algorithms}. In a third example, Buolamwini, a researcher with expertise in algorithmic bias and accountability, embarked on the Gender Shades project to formally audit facial recognition systems after facial analysis software that she needed to use for an engineering project would not recognize her face unless she wore a white mask \cite{AJLmission}. 

In other cases, everyday users may have little algorithmic expertise. For example, many small business owners collectively initiated an everyday audit of Yelp’s algorithm after noticing that positive reviews were not showing prominently on their business pages \cite{eslami2019user}. As another example, a group of Google Maps users uncovered that searching the N-word directed them to the White House, which Obama then occupied \cite{Fung2015map}. Since ``expertise’’ is multi-faceted, everyday users may have some relevant technical expertise, while lacking specific knowledge of machine learning.

\textbf{Collectiveness:} Everyday algorithm audits can also vary in the extent to which users work together collectively. Some cases are more individually led, such as the case when Alciné discovered that Google’s image recognition algorithm labelled some Black people as ``gorillas'' \cite{guynn2015google}. Similarly, Sweeney conducted her audit alone, a highly individual case \cite{sweeney2013discrimination}. Other cases, such as the time when LGBTQ+ YouTubers came together to investigate the demonetization of their videos, are more collective \cite{romano2019group}. It is worth noting that having a large number of users involved does not always mean an everyday audit is collective. For example, in the case where Booking.com users instigated an audit \cite{eslami2017careful}, they did so with little ability to communicate amongst themselves, resulting in a less collective audit despite its large scale.

\textbf{Organicness:} Though everyday algorithm auditing is based on everyday use, in practice the process can be more or less organic. Some everyday audits begin with an outside intervention but become more organic and user-led over time. For example, ImageNet Roulette was a simple online tool built by artists and researchers to support users in exploring and interrogating the input/output space of an image captioning model trained on the ImageNet dataset. However, after the tool was released to the public, users had autonomy in deciding how to use it. Collective auditing behaviors emerged organically through discussions on social media, as users shared findings and hypotheses and sometimes built upon each other’s explorations \cite{metz2019nerd,crawford2019excavating}. By contrast, other everyday audits start organically but later turn highly organized. For example, Sweeney, Noble, and Buolamwini all commenced their everyday audits inadvertently through everyday use, then turned to more formal auditing methods for testing and analysis \cite{sweeney2013discrimination,noble2018algorithms,buolamwini2016m}. Still other everyday audits are organic throughout the entire process. In the case of racial bias in Twitter’s cropping algorithms, Twitter users started and self-directed the whole audit with minimal intervention (in the form of a few tweets from Twitter employees whose input was invalidated by the users’ auditing process\footnote{\url{https://twitter.com/grhmc/status/1307435775856910343} \& \url{https://twitter.com/ZehanWang/status/1307461285811032066}}) \cite{mehta2021why}.

\subsection{Everyday Algorithmic Resistance}
We argue that everyday algorithm auditing can be thought of as a form of \emph{everyday algorithmic resistance} \cite{velkova2019algorithmic}. In their seminal work, Certeau and Scott \cite{de1984practice,scott1985weapons} developed the idea of “everyday resistance” to examine the ways in which people exercise their agency in front of dominant power structures and hegemonic culture formats in their everyday lives. Instead of thinking of resistance as organized actions that pose a revolutionary challenge, they foreground a different type of resistance that is less visible and more incidental but nevertheless constantly contesting the existing power relations in everyday forms. 

In the algorithmic era, we see similar resistance by everyday users around algorithmic systems. For example, in his analysis of gay men's resistance of Blued, China’s largest gay dating app, Wang looked how users act on the data they provide to dating apps to shape algorithmic dating outcomes in a different direction than what the system was originally designed for \cite{wang2020calculating}. Similarly, Devito et al. examined how Twitter users mobilized via hashtag to resist the introduction of algorithmic curation of Twitter's timeline \cite{devito2017algorithms}. Instead of characterizing the relationship between users and algorithms as passive consumers versus omnipotent machines, this line of research helps us foreground users' agency, capacity, and tactics in their regular interactions and contestations with algorithms. 

As a form of algorithmic resistance, in everyday algorithm auditing we see how users constantly test the limits of algorithms in their everyday use of the system. Sometimes, these auditing behaviors might be incidental and incremental: for example, a user might encounter harmfully biased algorithmic outputs and merely report their observations online. In other cases, these incidental and incremental behaviors might spur a group of users to work together, build on each other’s findings, and collectively interrogate the underlying system. 

\subsection{Counterpublics}
We argue that everyday algorithm auditing, when performed by a group of users collectively, can also be viewed as a form of \emph{counterpublics}. Nancy Fraser developed this influential idea as a critique to the concept of a universal public sphere developed by Habermas \cite{habermas1991structural}. She proposes that counterpublics can be understood as “parallel discursive arenas” \cite{fraser1990rethinking}, where members of often disadvantaged and marginalized social groups come together and participate in their own form of collective sensemaking, opinion formation, and consensus building. 

In the past a few years, we have seen growing efforts in building counterpublics around algorithmic systems. For example, Geiger examined bot-based collective blocklists in Twitter, which have been developed by volunteers to combat harassment in the social networking site \cite{geiger2016bot}. Xiao et al. \cite{xiao2020random} discussed how finstas\thinspace—\thinspace secondary Instagram accounts that feature content often unacceptable to users’ primary accounts\thinspace—\thinspace have developed forms of counterpublics for young people to share emotional or vulnerable content with their close friends. 

In everyday algorithm auditing, we saw similar counterpublics emerging. Indeed, many influential cases started with individual users complaining about certain socially harmful machine results in their everyday lives, but ended up with a large group of users working together for a collective and collaborative action. For example, in 2019 a group of YouTubers came together to show that the YouTube recommendation algorithm demonetizes LGBTQ+ content, resulting in a huge loss of advertising revenue for LGBTQ+ content creators \cite{romano2019group}. They tested the system with new examples, validated or invalidated each other’s findings, and interdependently probed the possible working mechanisms of the underlying system. During the process, they formed temporary counterpublics against the dominant algorithm, collectively making sense of the system, forming their own understanding and opinion, and building consensus toward possible remediations. In contrast to crowdworkers independently working on decomposed subtasks determined entirely by outside parties, in everyday algorithm auditing, the crowd collectively decides the course an audit takes, often through discussion threads or other social channels. For example, in the YouTube LGBTQ+ demonetization case, users started forming counterpublics on different platforms such as content creation websites (e.g., Patreon) and social media (e.g., Twitter) to work together detecting algorithmic biases as well as supporting affected communities. 

When forming counterpublics, users are involved in a process of \textit{collective sensemaking} \cite{weick1995sensemaking} of observed algorithmic behaviors. A rich tradition in HCI and CSCW has contributed to our understanding of how groups of people can work together, build upon each other’s explorations and insights, and work to achieve common goals. Past work has examined collective sensemaking across a range of domains, including citizen science \cite{cranshaw2011polymath}, knowledge mapping and curation \cite{halfaker2011don, geiger2013using, fisher2012distributed}, and social commerce \cite{chen2020understanding}, to name just a few. Similarly, in everyday algorithm audits, users come together to collectively question, detect, hypothesize, and theorize about problematic machine behaviors during their interactions with algorithmic systems.

\section{Methods}
We set out to understand what we can learn from existing everyday algorithm audits in order to inform the design of platforms and tools that can better support these practices. Thus, we asked: How can we better understand the characteristics, dynamics, and progression of everyday auditing practices? How can we support everyday users in detecting, reporting, and theorizing about problematic machine behaviors during the course of their interactions with algorithmic systems? 

In order to understand the nascent phenomenon of everyday algorithm auditing, we adopted an exploratory case study approach \cite{ragin1992case}, examining several recent, prominent cases. We began with a small set of high-profile cases that at least one of the co-authors was already familiar with. We iteratively reviewed and discussed these known cases to form the initial idea of ``everyday algorithm auditing’’. Through this process, we generated related keywords (e.g., “auditing,” “testing,” “algorithm,” “algorithmic,” “bias,” “harm,” “users,” “collective,” and “community”). Next, we searched news media (e.g., Google News, Google Search) and social media (e.g., Twitter, Facebook) using combinations of these keywords to get a rough sense for the scope of this phenomenon. The search yielded 15 cases (see Table 1) that met our definition for everyday algorithm auditing: one or more everyday users act to detect, understand, and/or interrogate biased and harmful algorithmic behaviors through their everyday use of a platform. These cases span multiple algorithmic domains, including image captioning, image search, machine translation, online rating/review, image cropping, credit scoring, and advertising and recommendation systems. We reference these cases throughout the paper to ground our discussion in concrete examples. However, this set of cases is by no means comprehensive. Furthermore, as discussed in our \emph{Design Implications} section, we expect that these existing cases represent only a thin slice of what is possible for everyday algorithm audits in the future. At this exploratory stage, our aim is to begin formalizing this concept, to help guide future empirical and design research in this space.

\textbf{Case Selection:} To support depth as well as breadth in our analysis of everyday algorithm audits, we chose a small subset of four cases to examine in greater detail. To select these cases, we iteratively extracted patterns from our initial dataset through a series of discussions among our research team. We set out to choose a set of cases that a) span multiple domains and vary along the three dimensions outlined in the \emph{Scope and Boundaries} section (algorithmic expertise, collectiveness, organicness) and b) were accessible to us via multiple data sources (e.g., user discussion posts, news articles, research studies), enabling a rich analysis. 

In particular, we chose two different domains that each have each been the target of everyday algorithm audits in recent years: image cropping algorithms and rating platform algorithms. Within each domain, we examined two cases that vary across the three dimensions discussed above to support comparison. Figure 1 summarizes the four cases, visualizing varying degrees along each dimension. For example, there are highly collaborative, less collaborative, and individual levels of collectiveness among the four cases. In addition, these cases span multiple levels of algorithmic expertise and organicness. The cases serve as illustrative examples to better understand how everyday audits can work, what paths they can take, and what impacts they can have. 

For each of our four primary cases, we gathered and reviewed discussion threads from the platforms where the everyday audits took place, then supplemented this information by drawing upon relevant media (e.g., news articles, academic publications) collected in our earlier search. We describe these four cases, their characteristics, and their dynamics in the next section. From this process of examining the progression of each case, a set of broader lifetime dynamics emerged, which we describe later in the paper.

\section{Case Studies}
Now, we focus on four cases of everyday algorithm auditing to illustrate and investigate the characteristics and dynamics of these audits more broadly. We look at two algorithmic sociotechnical platforms that each have been the target of everyday algorithm auditing in recent years. For each platform, we examine two different categories of biases that were detected via everyday algorithm auditing, as well as the different paths users took to conduct these audits. We compare these cases with each other to understand similarities and differences between everyday audits, and we consider how the paths taken contribute to their impacts.

\subsection{Twitter Image Cropping Algorithm}
Twitter employs algorithms throughout its infrastructure to make decisions about how content is presented to users. One such algorithm on Twitter uses neural networks to automatically crop the images users tweet, with the aim of improving users’ experiences \cite{theis2018neural}. Twitter crops images included in tweets based on how many images appear in a tweet, where more images contained in a tweet mean that each image appears as a smaller thumbnail; the dimensions for this cropping have been shared widely among Twitter users trying to understand why images in their tweets appear the way they do.\footnote{\url{https://twitter.com/nanobop/status/1255002567131557888} \& \url{https://twitter.com/kophing\_/status/1028000217654652928}} Twitter’s cropping algorithm attempts to find the parts of images most interesting to people\thinspace—\thinspace as judged by how likely people are to look at a certain region and termed “saliency”\thinspace—\thinspace and center that part of the image within the frame, theoretically cropping the least interesting parts of the image out of the thumbnail \cite{theis2018neural}. This appears to be a good solution for users and Twitter alike: maintain consistent design of tweets while highlighting the part of an image users most want to see.

\begin{figure}
\begin{centering}
\includegraphics[scale=0.28]{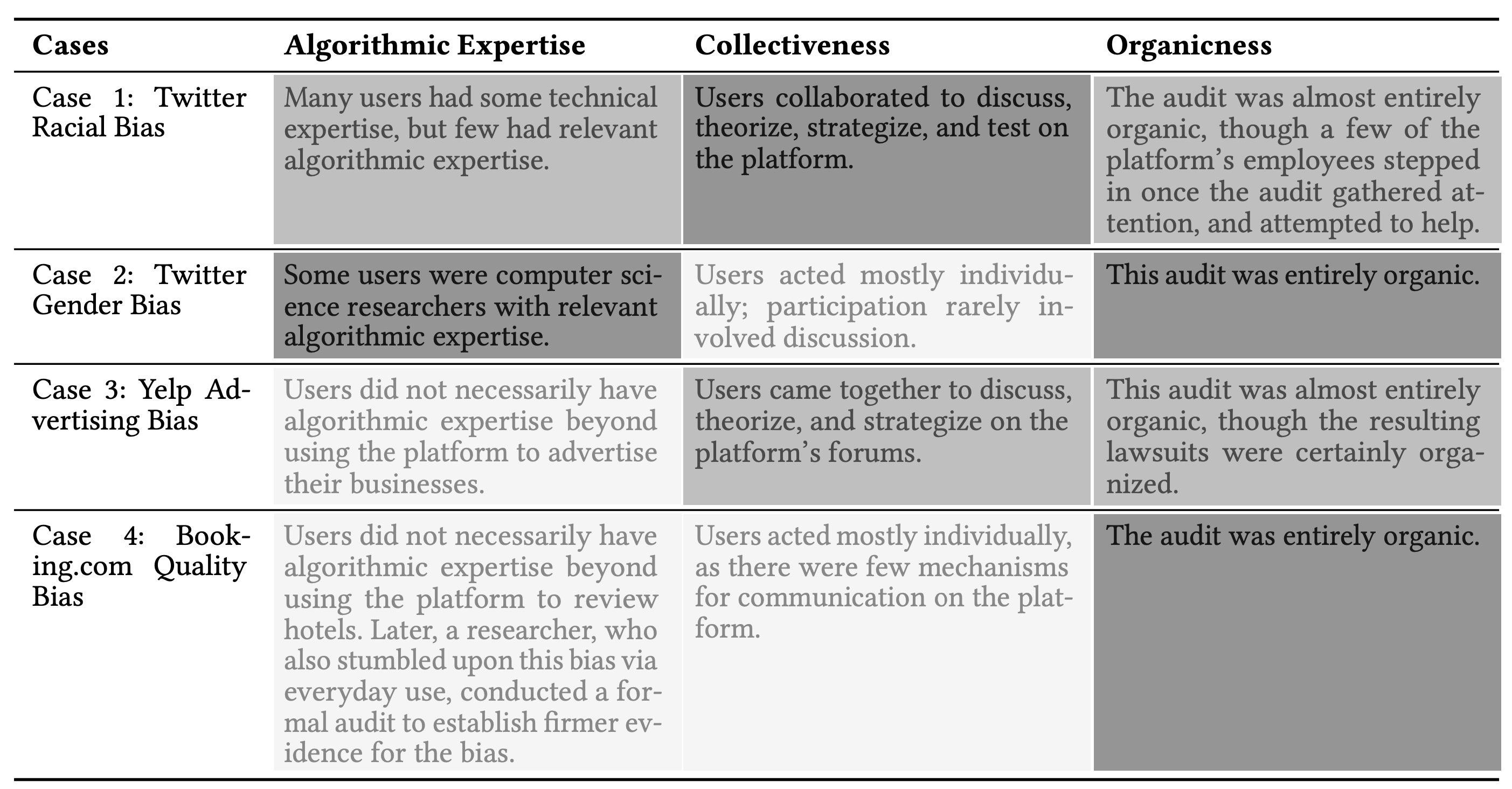}
\caption{High-level descriptions of our four primary case studies across the three dimensions outlined in the \emph{Scope and Boundaries} section. These cases serve as illustrative examples to better understand how everyday audits can work, what paths they can take, and what impacts they can have. Darker shades represent higher levels along a given dimension; lighter shades represent lower. Note that less variation is presented along the ``Organicness'' dimension given that this paper focuses on everyday audits toward the more organic end of the spectrum as less organic everyday audits have been well described by past literature. }
\label{tab:4cases}
\end{centering}
\end{figure}

But Twitter users noticed that the image cropping did not always match what they would have guessed to be the most “salient” parts of images: specifically, they noticed problematic trends in the behavior of the automated cropping, leading them to suspect bias within the cropping algorithm. In one high-profile case in September 2020, Twitter users investigated potential racial bias in the way the cropping algorithm decided whom to focus on and whom to crop out. In another case that gained less traction in January 2019, Twitter users considered possible gender bias in the cropping algorithm’s decisions. Below, we describe these two cases and the paths they took along with their dynamics, results, and impacts.

\subsubsection{Case 1. Racial Bias:} Our first case began when a user of the video conferencing platform Zoom realized that a colleague’s trouble using virtual backgrounds might have to do with the software not recognizing darker skin, then tweeted about it.\footnote{\url{https://twitter.com/colinmadland/status/1307111816250748933}} The user tweeted images with the explanation, single shots that captured the Zoom screens of both the colleague, who has darker skin, and the tweeting user, who has lighter skin, side-by-side. The final tweet cropped this wide image to show a thumbnail of just the lighter-skinned person, prompting the tweeter to notice unexpected behavior and theorize that Twitter’s cropping algorithm might be racially biased, then tweet about this new subject.\footnote{\url{https://twitter.com/colinmadland/status/1307115534383710208}} The individual tweeting\thinspace—\thinspace that is, the initiator of the everyday audit\thinspace—\thinspace stumbled upon Twitter’s photo cropping issue in the midst of regular use of Twitter: sharing information with others on the platform. Incidentally, the information the initiator set out to share on Twitter was related to the issue noticed on Twitter itself, which may have helped the user enter the correct headspace for more awareness of these types of strange and potentially biased behaviors.

Having developed a hypothesis that the cropping issues were related to skin color, the initiator kicked off testing by tweeting the same image as before but flipped so that the person initially on the right was now on the left and vice versa.\footnote{\url{https://twitter.com/colinmadland/status/1307130447671984129}} This test also showcased the lighter-skinned person. Others on Twitter began to join in, tweeting images of their own creation to see what the algorithm would prioritize and what the algorithm would crop out. At the same time as users tested whether the algorithm was indeed biased by skin tone, other users developed new hypotheses that would lead to more tests in attempts to confirm or invalidate. 

For example, one highly shared thread put Mitch McConnell at one end of a long image and Barack Obama at the other, finding that McConnell appeared in the thumbnail no matter which end he appeared on.\footnote{\url{https://twitter.com/bascule/status/1307440596668182528}} When a new hypothesis surfaced suggesting that the preference in this case might be based on the clothing, specifically the ties, that the two politicians wore in the photos used\thinspace—\thinspace perhaps the algorithm prioritizes the color red, which is worn by McConnell while Obama wore a blue tie\thinspace—\thinspace a new test was run: exactly the same as before, but switching the ties worn by the two men. McConnell again appeared in both thumbnails. 

Other theories arose, leading many users to conduct a wide range of tests in tweets involving elements such as background shade,\footnote{\url{https://twitter.com/grhmc/status/1307435994246008844} \& \url{https://twitter.com/m\_paardekoper/status/1307636655508140032}} background color,\footnote{\url{https://twitter.com/kosmar/status/1307777202445004800}} cartoon characters,\footnote{\url{https://twitter.com/RasmusMalver/status/1307615213810839552} \& \url{https://twitter.com/\_jsimonovski/status/1307542747197239296}} men,\footnote{\url{https://twitter.com/IDoTheThinking/status/1307505161640660992}} women,\footnote{\url{https://twitter.com/IDoTheThinking/status/1307449013247991808}} different numbers of people,\footnote{\url{https://twitter.com/onyowalkman/status/1308035295095316481}} a variety of skin tones,\footnote{\url{https://twitter.com/joyannboyce/status/1308080366369021954} \& \url{https://twitter.com/kosmar/status/1307776710604185600}} image contrast,\footnote{\url{https://twitter.com/m\_paardekoper/status/1307636653683601408} \& \url{https://twitter.com/m\_paardekoper/status/1307641363756965891}} and even dogs.\footnote{\url{https://twitter.com/ameliorate\_d/status/1307576187942703104} \& \url{https://twitter.com/MarkEMarkAU/status/1307616892551487488}} Throughout, users built on and countered each other’s suppositions, wielding the tests they had run as evidence supporting or opposing the claims of various hypotheses. 

The audit by everyday Twitter users of Twitter’s cropping algorithm for racial bias had both specific and broader impacts. Broadly, the highly collective nature of this audit where users worked together to determine a direction for the audit, constantly interacting with each other along the way, led to the creation of a large counterpublic space in which thousands of Twitter users interacted.\footnote{e.g., \url{https://twitter.com/bascule/status/1307440596668182528}} This space allowed users to question the algorithmic authority of the very platform they conversed on, develop concrete concerns, and push for change. As might be expected for a group this large, the media picked up on what was happening and many articles were published about the issues highlighted by the audit (e.g., \cite{johnson2020apparent,bbc2020twitter,brooks2020twitter,sanjay2020twitter}). In addition to\thinspace—\thinspace or perhaps because of\thinspace—\thinspace these broad reverberations, Twitter provided an official response addressing how they would attempt to fix the issues highlighted by the everyday audit: ultimately, they aimed to reduce cropping’s reliance on an ML model and allow for more user controls as well as ensure that image thumbnails in the preview of a tweet accurately represented what the image thumbnails would look like once the tweet was live \cite{Twitter2020}.

\subsubsection{Case 2. Gender Bias:} The second case began when VentureBeat shared an article with an image of two women and two men on Twitter.\footnote{\url{https://twitter.com/VentureBeat/status/1080485485473083392}} A Twitter user pointed out that both women’s heads had been cropped out while neither of the men’s had. As a result, while the cropping algorithm highlighted the men's heads, it focused instead on women's chests.\footnote{\url{https://twitter.com/mraginsky/status/1080568937656565761}} The individual who shared this issue to start the everyday audit had noticed it while looking at the VentureBeat tweet, a normal usage of the Twitter platform. Coincidentally\thinspace—\thinspace ironically?\thinspace—\thinspace the VentureBeat article featured four prominent AI researchers, with one focusing on responsible AI \cite{johnson2019ai}, so this may have nudged the user into a mindset that took notice of the cropping issue.

Unlike the previous, racial cropping issue, very few other Twitter users joined this everyday audit. Because of this, there was little additional hypothesis forming or testing, though some conversation did happen in replies to the initiating tweet\footnote{\url{https://twitter.com/mraginsky/status/1080569326644748289}} as well as in a thread begun by another user\footnote{\url{https://twitter.com/timnitGebru/status/1080697598170808321}} that mostly focused on whether Twitter or VentureBeat was to blame. The gendered cropping issue was revisited in September 2020 and received more (but still relatively little) engagement, including minimal testing of how Twitter crops other images, on the heels of the prominent racial cropping case discussed above.

Overall, the gendered cropping issue gained little traction. It stayed more individualized: only a handful of people participated when it was first brought up in 2019, and though over a thousand liked the 2020 tweet resurfacing the issue, still very few engaged in any sort of conversation about the issue. One might suspect that the initiators of the gender bias case were not as influential or active as the initiators of the racial bias case; but the initiators and of both the racial cropping audit and the gender cropping audit have comparably large Twitter followings. Examining this under the lens of meme virality produces some potential explanations for why this might have occurred. While both Twitter cases can be understood as a meme in the sense that it is a “unit of cultural transmission” \cite{dawkins2016selfish}, this one lacks the elements of mimicry and remix some view as vital to meme-hood and virality \cite{shifman2013memes}. In the previous case of racially biased cropping, mimicry and remix can be seen as users build upon each other’s observations, hypotheses, and testing strategies in the form of tweets and retweets. Another possible explanation for the disparity in engagement is that current events in 2020, such as the ubiquity of the Black Lives Matter movement, might have led people to be more aware of and more interested in issues of racial justice. This is a particularly appealing possibility given that information in memes tends to propagate from mainstream news to social media \cite{leskovec2009meme}. Since content with greater emotive strength is more likely to spread \cite{guadagno2013what}, another possibility is that Twitter users are less interested in gender-related bias than race-related bias. However, in November 2020, a Twitter user apparently unconnected to either of the past conversations around gender cropping tested gender bias in a thread that received almost four times the amount of engagement as gender cropping before, suggesting that people are interested in the gender cropping issue as well.\footnote{\url{https://twitter.com/dikili/status/1326674332135854080}} So, why did these two types of everyday audits, both occurring on the same platform due to biases introduced by the same algorithm, receive different levels of attention and take diverging paths? This remains an open question that needs further investigation. 

Likely because of the small number of people involved in the gender cropping audit, it had little impact. A small counterpublic space did arise, briefly, but not much else occurred. The small number of people kept the audit mostly individualized in nature, with people who engaged mostly supporting the audit’s existence rather than participating or helping to direct it. Not many people meant that the audit had much more difficulty gaining any sort of traction beyond itself and was ultimately unable to garner publicity or platform change like the racial cropping audit.

\subsection{Rating Platform Algorithms}
Rating platforms are another type of sociotechnical system that use algorithms to determine how to present information to users. Final ratings can have huge impacts on a business’s livelihood as consumers look to the ratings to direct their behaviors: for instance, restaurants were 30–49\% more likely to fill their seats following just a half-star rating increase on Yelp \cite{anderson2012learning}. Though some users might assume that a final rating is determined as a simple, raw average of all the ratings given by individual users, many rating platforms do not follow this method. For example, in calculating a product’s rating, Amazon considers factors including how old the associated review is, whether the reviews have verified purchases, and the number of votes the review received for being helpful \cite{bishop2015amazon}. Other platforms like Yelp \cite{yelp} do follow this method of straightforward averaging but only include authentic reviews, as classified by their algorithms. 

Additionally, the mechanisms behind these rating systems often lack transparency \cite{eslami2019user}, which has caused controversy about algorithmic bias in recent years. In May 2016, for example, Uber was accused of exploiting its rating system’s opacity to slowly decrease the ratings of its Australian drivers so that it could suspend them and charge high reinstatement fees \cite{tucker2016uber}. Similarly, Yelp has been criticized for lack of transparency around the inner workings of its rating system \cite{fowler2011judge}. After suspecting that the platform manipulated its rating system in order to force businesses to pay for advertising, Yelp users conducted an investigation. Below, we explore the Yelp everyday algorithm auditing case as well as the Booking.com case in which users scrutinized the rating algorithm after realizing the ratings appeared mismatched with their expectations. 

\subsubsection{Case 3. Yelp Advertising Bias:} The Yelp case has a long history, beginning as early as 2011 when small business owners on Yelp began to suspect that Yelp manipulated its review filtering algorithm to hide positive reviews,\footnote{\url{https://www.yelp.com/topic/los-angeles-very-bothered-by-yelp-review-filter-part-1}} which would then compel business owners to pay for advertising to boost their rating positions to make up for having fewer positive reviews. Many small business owners on Yelp claimed to have noticed this apparent algorithmic aberration after they had received phone calls from the Yelp sales team about advertising with Yelp; when they rejected the offer, opting not to pay for Yelp advertising, the positive reviews for their businesses started to get filtered out. Even worse, the negative reviews for their businesses inexplicably moved to the recommended page. The business owners took, naturally, to Yelp, where they began posting on the Yelp Talk discussion forums about what they had noticed.\footnote{e.g., \url{https://www.yelp.com/topic/los-angeles-very-bothered-by-yelp-review-filter-part-1}} In this way, the business owners together initiated the everyday algorithm audit by sharing the perceived issue with others after encountering it through typical usage of the platform. 

Having developed a hypothesis that the filtering issues were related to the advertising offer, business owners began to test by seeing what happened to reviews for businesses following calls from Yelp Sales agents.\footnote{e.g., \url{https://www.yelp.com/topic/miami-yelps-filtering-seems-like-extortion}} Results of these tests were circulated to others via the discussion forums, where business reviewers joined business owners in conversation about the filtering algorithm. Alternate hypotheses arose, such as the theory that reviews were being filtered out because writers appeared illegitimate based on profile appearance or behavior\thinspace—\thinspace e.g., no profile picture and few friends; writing too few, too pithy, too extreme, or too infrequent reviews.\footnote{\url{https://www.yelp.com/topic/denver-yelp-routinely-filters-my-customer-reviews-because-i-wont-advertise-with-them} \& \url{https://www.yelp.com/topic/walden-why-does-my-yelp-review-not-show-up-on-the-company-site}} These conjectures were often based on users’ folk theories about how the filtering algorithm operated and not on concrete data \cite{eslami2019user}. Yelp users probed these hypotheses through debate in the forum and through further testing that often involved business owners analyzing the elements of filtered and recommended reviews and business reviewers visiting establishments and leaving reviews to test whether they would be visible.\footnote{\url{https://www.yelp.com/topic/dallas-yelp-filtering-almost-all-reviews-removing-more-by-the-day}} Business owners participated especially intentionally, as they often had personal motivation if the algorithm had started filtering their business’s positive reviews or recommending their business’s negative reviews. Throughout, the users referred to the tests they had run and witnessed as evidence in support of or in opposition to various proposed hypotheses. 

The audit of Yelp’s review filtering algorithm by everyday Yelp users has a number of ramifications. In perhaps the largest and most visible result, the claims of the audit led to many lawsuits filed over the years with almost 700 accumulated over time \cite{kang2013yelp,palmiotto2020yelp}. Though these lawsuits were ultimately unsuccessful \cite{goldman2014yelp}, they created a large amount of public awareness and forced Yelp to engage with the complaints, at least in the legal arena. Perhaps the widespread initiation of this everyday audit, with different business owners in widely separated locations noticing the inconsistency in the filtering algorithm’s behavior at the same time in a sort of collective initiation, forged the way for such large impact. In a smaller but still important way, the everyday audit created awareness and support in a counterpublic space within the Yelp platform on Yelp Talk discussion forums as users collectively considered the filtering algorithm. Yelp unintentionally allowed this, but it formed an imperative aspect of this everyday audit and highlights the value of providing an ecosystem for communication and discussion between audit participants.

\subsubsection{Case 4. Booking.com Quality Bias:} The other case of rating platform bias began when Booking.com users rated hotels where they had stayed and noticed a discrepancy between their desired total rating and what the algorithm calculated. Booking.com asks users to assess a number of different elements of their lodging experience, then uses an algorithm to calculate an overall rating based on this assessment. The discrepancy appeared when disgruntled hotel guests selected the lowest option for every element, aiming for a score of 1 or 0 out of 10, only to discover that the algorithm would produce a score of 2.5 \cite{eslami2017careful}. Similar to the previous, Yelp case, many users realized this strange Booking.com behavior around the same time, initiating the audit by pointing out the issue in the body of their hotel reviews, and they did so during typical platform usage: trying to leave poor reviews. 

Unlike the Yelp case, this everyday algorithm audit stayed highly individualized. After Booking.com users reported the issue they observed through reviews, there was no on-platform way for the users to communicate with each other, in contrast to the Yelp case where users could talk on the Yelp platform. Booking.com users thus individually set out to test their hypothesis that the rating algorithm would not dip below 2.5 by trying different combinations of rating inputs. Multiple hypotheses for the Booking.com behavior eventually did emerge, such as that the algorithm was purposefully skewing the ratings to present a better image for low-to-medium quality hotels, that the algorithm rated overly high in general, and that the algorithm was just incorrect \cite{eslami2017careful}. Users attempted to warn others of both the bad hotel experience and the unrepresentative algorithm behaviors, using review writing as a method for raising awareness since this was the only way to communicate on the platform. 

In terms of impact, this everyday audit had few. The Booking.com case generated little publicity, despite having proven existing bias. This is especially interesting when taken in conjunction with the Yelp case, which generated an abundance of publicity despite never having proven bias\thinspace—\thinspace in fact, the Yelp case was ultimately dismissed due to the lack of scientific evidence \cite{goldman2014yelp}. Perhaps this difference arises from a difference in personal injury: in the Yelp case, the livelihoods of many business-owning everyday auditors were at stake, whereas in the Booking.com case, the most at stake for the everyday auditors was the ability to fully trust the ratings upon which they based hotel reservation decisions. Or perhaps more collective everyday audits invite publicity, as the collective nature means that not only are more people aware of the audit occurring, but also they are more aware of it as they actively participate in sensemaking discussions and actions. While Booking.com’s audit participants lacked collectiveness, they still changed their behaviors to try to make changes to the system; that is, they signaled to other users a more accurate presentation of their hotel stays, and they trusted the overall platform ratings less \cite{eslami2017careful}. Booking.com did eventually make changes to its rating system to make the lowest possible score to 1 (instead of 2.5) and allow users to select an overall score that will appear unchanged by an algorithm \cite{brian2019booking}, which are likely indirect results of this everyday audit.

\section{The Lifetime and Dynamics of an Everyday Audit}
Below we propose a process-oriented view of everyday algorithm audits, with phases ranging from (1) initiation of an audit to (2) raising awareness of observed issues to (3) hypothesizing about observed behaviors and testing an algorithmic system to (4) some form of remediation. In practice audits may follow a non-linear path through these phases, and may terminate before reaching all phases, for various reasons discussed below. Regardless, these provide a useful high-level understanding for describing the paths everyday audits can take, comparing different everyday audits, and envisioning interventions to facilitate such audits.

\subsection{Initiation}
Everyday algorithm auditing usually starts in the moment that an individual or group of individuals finds an instance of harmfully biased behaviors through normal usage of an algorithm. Some of the cases such as the Twitter cropping algorithm’s racial and gender bias start with one individual user, the \emph{initiator}, and then continue via other users. On the other hand, in some cases such as Yelp's filtering algorithm or Booking.com’s rating algorithm, many users notice a potential bias around the same time in different locations with or without previous connections with each other. These behaviors, as we call individual and collective initiation respectively, could play a role in the spread and impacts of an everyday audit. 

Everyday audits are often initiated by \emph{incidental} exposure to problematic machine behaviors: in other words, users may not necessarily intend to conduct an audit from the start. Rather, users often stumble upon concerning algorithmic behaviors in the midst of their day-to-day use of a system. For example, in the Twitter racial cropping case began when, in the course of tweeting an image, noticed that out of two people shown in the image, the person with darker skin was excluded from preview thumbnails. While everyday audits frequently begin through incidental exposure, users may be more likely to detect and spread awareness of problematic machine behaviors when they are already “primed” to look for such behaviors. For instance, in the Twitter racial cropping case, the audit initiator noticed a concerning cropping behavior in the context of tweeting about a different racial bias observed on another platform (Zoom). As might be expected, unintentional initiation is more likely to arise from the users’ situated actions whereas more intentional initiation is more likely to arise from the actions of external auditors and not in the context of everyday use.

\subsection{Awareness Raising}
After detecting a problematic algorithmic behavior, the initiator or initiators broadcast what has been discovered to others. For example, in both Twitter cropping cases, this took the form of a tweet to other users. Others can then spread the word to even more people. Promoting the audit is a valuable part of the process, as it can bring more people into the discussion and increases visibility of the issue at hand. Sometimes, like in the racial cropping Twitter case, this promotion by other users creates the opportunity for further sharing and for participating in other parts of the audit like hypothesis forming and testing, which we explain in the next section. Other times, when discussion between users is more limited by a platform like in the Booking.com case where users had to communicate through reviews, this promotion by other users solely creates more awareness among those who see it, helping them to adjust their own behavior in using the system.

Raising awareness has ramifications both internal and external to the everyday algorithm audit. Internally, sharing information about the audit, for example in the racial cropping case on Twitter when some people created larger threads combining information from multiple places, serves to draw more participants into the audit in a cascading, self-referential way that potentially increases its collective nature. That is, the more users promote the audit, the more new people notice its happening and are then more likely to promote the audit themselves. Externally, sharing information about the audit makes it more likely that the awareness will eventually extend beyond the audit, as will be discussed more in Remediation below. 

Everyday audit participants also can generate awareness through supporting the mechanisms of the audit. The LGBTQ+ YouTube demonetization is an example of this, as some YouTubers worked together to collect keywords, and a large number of people did a lot of YouTube tagging as well as shaming the company on Twitter for its actions. While this can promote the audit itself, these actions are more focused on assisting\thinspace—\thinspace but not guiding\thinspace—\thinspace other parts of the audit. Often a core group of people collaborate in the auditing process and are surrounded by a larger ring of people supporting their actions. The lack of many people in this support ring might help explain the lower traction for the Twitter gender cropping case as compared to the racial cropping case. Both the core group and the surrounding ring form a counterpublic space that supports the everyday audit as it progresses.

\subsection{Hypothesizing \& Testing}
Following initiation and/or awareness raising, an everyday algorithm audit may progress toward users forming hypotheses for the observed behavior and intentionally testing an algorithmic system to collect further evidence. This process may involve just the individual who initiated the audit or may extend to many users if the initiator raised awareness and inspired others to get involved. Individual users who proceed alone continue with little support either from other situated users or from external experts. In more collective everyday algorithm audits, a group of everyday users organizes and builds on each others’ efforts in the process. The users collectively decide on and steer the process, often through online discussion. 

When users start hypothesizing about the potential biases of a system and testing their hypotheses, they usually develop “folk theories”: non-authoritative theories a user forms to explain and understand how a technological system works \cite{eslami2016first,devito2018people}. These theories, regardless of their validity, shape user behaviors and inform how they act around a system. In an everyday algorithm audit, these theories help users to form hypotheses about the sources of a bias, then test them. Sometimes these tests fail, such as switching the tie colors of McConnell and Obama in the Twitter racial cropping algorithm case to test the hypothesis that the algorithm might prioritize red over blue;\footnote{\url{https://twitter.com/bascule/status/1307440596668182528}} other times these tests succeed, such as manipulating the algorithm inputs in the Booking.com case to test the hypothesis that the algorithm skewed the minimum score up \cite{eslami2017careful}. These examples show the importance of folk theory development in the process of an everyday audit: the more the platform allows users to develop and test theories, the more users have the opportunity to discover biases. One way to empower users in this process is \emph{seamful} design; as Eslami et. al. described, adding visible hints disclosing aspects of an automation process, or seams, into the design of an algorithmic system can help users construct better conceptual understandings and theories about the system \cite{eslami2016first}. This increased transparency can provide users with more visibility, which may aid in detecting potential biases. The increasing number of everyday algorithm audits, along with new regulations around algorithmic bias \cite{goodman2016eu,house2016big}, can push algorithmic sociotechnical systems for a more transparent and seamful design. 

Hypothesizing and testing often involves different degrees and forms of collaboration. In cases where users perform the auditing in an individual endeavor, they might spot an issue and quickly report it via social media channels or the platforms’ internal discussion forum or, if coming from a background of high technical expertise, they might conduct the auditing in a more systematic way alone. In other more collective cases, users come together to make sense of how the system actually produces such biased results. During this process, they might raise new evidence, test each other’s evidence, and collectively search for patterns that emerged from their discussion. For example, in Yelp's filtering algorithm case, users collectively worked together on the Yelp platform itself to form different hypotheses about how the algorithm works, test those hypotheses, and report the results back to the community. 

Similar to the motivations behind initiation of the audit, the hypothesizing and testing process can range from organized to organic. More organized testing often involves external stakeholders (e.g., researchers, developers, journalists) who intervene in the process, while more organic testing often arises naturally without outside intervention as everyday users detect and assess algorithmic behaviors and play an active role in determining the course the audit takes. We argue that everyday algorithm auditing can receive\thinspace—\thinspace and often benefit from\thinspace—\thinspace different degrees and forms of external intervention. For instance, in some cases, developers with insight into the inner workings of a given system may be able to provide useful feedback to guide hypothesis formation, such as in the Twitter cropping case when an engineering team lead chimed in to let audit participants know that facial recognition is not part of the cropping algorithm.\footnote{\url{https://twitter.com/ZehanWang/status/1307461285811032066}} To the extent that external interventions represent \emph{guidance} for an otherwise collectively-led audit, rather than directing or orchestrating the collective’s activities as in a crowdsourced/collective audit, we do not view such interventions as invalidating the \emph{everyday} nature of the audit.

\subsection{Remediation}
At its core, an everyday algorithm audit has a singular objective: instigate change based on the issues identified and investigated. Change can occur at multiple levels. One type of change creates or increases external awareness through publicity. As the audit gains traction, news and other media sources are likely to become aware of the audit as well and may publish articles on the audit and the issues it highlights. This leads to broad social impact via greater awareness of the issues in the public sphere by all types of people, who may then choose to share this information with still others, alter their own patterns of behavior, or even take part in the audit if it is still occurring. 

Another type of change comes in the form of legal action. For example, in the Yelp case, close to 700 lawsuits were filed in attempts to force Yelp to change its review filtering algorithm \cite{Kang2013}. Though none of the lawsuits in the Yelp example were successful, they had the potential to put the platform in a position in which it would have been required to respond and enact change or risk behaving illegally. There are also less direct repercussions of the legal action. Regardless of whether a lawsuit has the desired result for the plaintiff, the filing of a legal case in itself creates publicity for the issue. 

Perhaps the most effective type of change is at the platform level. Sometimes when an everyday algorithm audit brings a problem to the forefront, the creators or employers of the algorithm take notice and agree that a change should occur. An official statement can create legitimacy around the audit and the identified issue, and swift redress can occur, since those who have decided on change now include those who have the power to enact change. Both of these resulted from the Twitter racial cropping case, as Twitter posted on the company blog \cite{Twitter2020} and committed to changing the way that image previews are cropped. In another example, Google revised its image recognition algorithm’s racial bias in classifying some Black users as gorillas (which was found and reported by everyday users) \cite{vincent2018google}. However, platforms’ attempts at redress are not always effective. For example, Google preemptively announced in 2020 that it had fixed its translation classifier to address bias that assigned gendered pronouns in an often stereotypical way to professions and activities when translating gender-neutral languages \cite{olson2018algorithm, wiggers2020google}; the issue reemerged in 2021 with other languages.\footnote{e.g., \url{https://twitter.com/frostaf/status/1376699332917870592}, \url{https://twitter.com/imajeanpeace/status/1374135940898156544}, \& \url{https://twitter.com/DoraVargha/status/1373211762108076034}}

When platforms fail to act to fix a harmful and biased algorithmic behavior, everyday users sometimes go beyond detecting, testing, and reporting a bias to begin rectifying the behavior themselves. For example, in the Booking.com case, users started manipulating the algorithm to return lower algorithmically calculated ratings in their hotel reviews so that the actual ratings better aligned with users’ intended ratings \cite{eslami2017careful}. Everyday users also engage in what we call “collective repair” to resist and ameliorate the harm an algorithm can bring to a community. For example, in an everyday audit TikTok users found that the platform’s For You Page algorithm suppresses content created by people with certain social identities, such as LGBTQ+ users and people of color. In response, some users worked together to amplify the suppressed videos by specifically engaging with them and adding comments \cite{KARIZAT2021TikTok,simpson2021you}. 

\section{Design Implications}
We have discussed users’ auditing behaviors around the Twitter image cropping algorithm and around rating algorithms on Yelp and Booking.com. Across these cases of everyday algorithm auditing, we observed similar patterns: a group of day-to-day users comes together to detect, understand, and interrogate problematic machine behaviors during the course of their ordinary interactions with algorithmic systems. In this section, we explore what lessons might be drawn from the study of everyday algorithm auditing to overcome limitations of previously proposed auditing approaches. We ask how designers might facilitate everyday algorithm auditing practices, building upon regular users’ existing motivations to engage in such auditing behaviors and supporting their unique strengths in surfacing harmful algorithmic behaviors. How might designers support and scaffold everyday algorithm audits, without intervening so heavily as to sacrifice their bottom-up, user-driven nature?

Below we outline five broad categories of potential design interventions\thinspace—\thinspace (a) community guidance, (b) expert guidance, (c) algorithmic guidance, (d) organizational guidance, and (e) incentivization\thinspace—\thinspace and discuss potential design trade-offs. 

\subsection{Design for Community Guidance} 
Across our cases, we observed different degrees and forms of collaboration among everyday auditors. In general, more collective efforts may lead to more impactful everyday algorithm audits. For example, everyday auditors in the Yelp case engaged in more collaborative auditing behaviors compared with the Booking.com case. We could speculate that, through increased interaction with other everyday auditors, Yelp users may have been more likely to form “counterpublics” against the dominant algorithmic system, in a collective effort toward sensemaking and consensus building. Moreover, more collective everyday auditing efforts may be more successful in raising awareness and pushing for changes on the algorithmic, platform, and even societal level, as we observed in our discussions around Twitter’s cropping algorithm. 

One way designers might support collective auditing behavior is to offer mechanisms that help everyday auditors guide each other’s efforts. At minimum, this might include providing spaces for community discussion so that everyday auditors can discuss, raise awareness about, and build upon each other’s findings, as we have already observed in the Yelp and Twitter cases. Apart from directly facilitating the auditing process, public discussion forums may help garner support from advocacy groups and news organizations to raise awareness and push for potential remediations on different levels, as we observed in the YouTube LGBTQ+ demonetization case. 

Beyond simply providing spaces for discussion, we might imagine tailoring or augmenting discussion channels with mechanisms that are explicitly designed to support everyday auditing. For example, a discussion platform designed for everyday auditing might allow community members to upvote specific algorithmic behaviors that other community members have reported, in order to collectively surface the most \emph{severe} reports, or reports that may \emph{benefit from further discussion}. Furthermore, when an everyday auditor selects a particular report, the discussion platform might provide an overview of relevant hypotheses and findings that other auditors have generated so far, to help the auditor build upon previous contributions instead of retreading old ground.

When designing to support community guidance, we should also consider the possibilities of leveraging existing platforms rather than creating entirely new ones. In the Yelp case, users were able to leverage embedded discussion forums to organize their everyday auditing. In the Twitter case, the platform being audited was itself a discussion platform, allowing users to discuss and test algorithmic behavior \emph{simultaneously}. This evokes Grevet and Gilbert’s “piggyback prototyping” technique in which existing, popular platforms (e.g., Facebook and Twitter) are appropriated in the design of new social computing systems in an attempt to help overcome the challenges of obtaining critical mass on entirely new platforms \cite{grevet2015piggyback}. A potential trade-off is that hosting community discussion \emph{within} the same platforms that are under audit may lead to concerns about platform-driven censorship or suppression, given uneven power dynamics and the potential for value conflicts between justice-oriented audits and profit-oriented platforms. We anticipate that in certain cases, everyday auditors may feel more comfortable hosting these discussions elsewhere.

\subsection{Design for Expert Guidance} 
Although the cases we reviewed were often initiated organically by situated users, others with relevant technical expertise sometimes weighed in as well. In the Twitter photo cropping case, developers from Twitter intervened in the discussion of racial bias to weigh in with some of the background knowledge they had available on how the cropping algorithm functions.\footnote{\url{https://twitter.com/ZehanWang/status/1307461285811032066}} We expect that everyday algorithm audits can benefit from expert input, whether from system developers, researchers, or regulators. However, it remains an open question what forms and degrees of expert engagement may be most helpful in facilitating everyday audits.

To help steer everyday auditors' ongoing efforts in more productive directions, designers might invite relevant experts to continuously monitor community discussions and intervene as needed \cite{chan2016improving,dow2012shepherding}. For example, at any stage of an everyday audit, a product team might weigh in and provide feedback, based on technical knowledge of their products, regarding the plausibility of user-generated hypotheses. However, if this is not handled carefully, experts’ involvement may risk diminishing everyday auditors’ sense of community and autonomy. For example, the insertion of technical or domain experts into the middle of an everyday audit might create unequal power dynamics in community discussions. 

Relevant experts may also provide guidance to users at the initiation stage. For example, to complement their own internal auditing efforts, a product team may actively draw users’ attention to suspected issues or components of a system that they would like users to audit, then suggest testing strategies for users to try \cite{holstein2019improving}. Similarly, in the case of ImageNet Roulette \cite{crawford2019excavating}, researchers built an interface to help everyday users quickly and easily audit a model trained on the ImageNet dataset, then disseminated this interface on social media. In turn, the ImageNet Roulette interface went viral, spurring a large and diverse group of users to test for algorithmic biases using images of their choosing. Although audits were initiated by technical experts in both of the above examples, users still had agency and autonomy in directing their collective efforts via social media. 

The design of bidirectional feedback mechanisms between developers and everyday users of algorithmic systems represents a critical direction for future research. In addition to the need for systems that help developers provide effective feedback and guidance to users who are engaged in everyday auditing behaviors, there are opportunities to design systems that scaffold users in providing more useful, actionable feedback to developers.

\subsection{Design for Algorithmic Guidance} 
For certain auditing tasks, such as exploring a system’s behavior, gathering evidence, or testing hypotheses, everyday auditors may benefit from algorithmic assistance. For instance, designers might develop auditing interfaces that automatically surface potentially important instances for everyday auditors to examine further. As a foundation for such interfaces, it may be possible to build upon emerging algorithmic techniques for crowd-in-the-loop detection of “unknown unknowns” in ML models (e.g., \cite{bansal2018coverage, Cabrera2021Deblinder, lakkaraju2017identifying, liu2020towards, ni2017reducing, vandenhof2019contradict}), which automatically surface cases that are more likely to be mislabelled and/or misclassified. These methods focus on surfacing regions of a model’s error space in which the model is highly confident yet incorrect \cite{lakkaraju2017identifying}. This form of algorithmic guidance could supplement guidance that users may receive from each other (e.g., via upvoting mechanisms) and from expert guidance (e.g., suggestions for how everyday auditors should direct their search efforts), potentially surfacing cases that these other forms of guidance would miss.

In some of the everyday auditing cases we have reviewed in this paper, establishing the presence of certain issues requires some form of \emph{comparison}. We distinguish between two major categories of auditing behaviors: “instance-based” and “comparison-based” auditing. Instance-based auditing applies to cases where the observation of an isolated harmful algorithmic behavior is sufficient to ground an argument for remediation. For example, when users found that the Google Photos app assigned a photo of Black people with the label “gorillas” \cite{guynn2015google}, it was not necessary to demonstrate that most photos of Black people receive the same label, nor was it necessary to demonstrate that Black people are consistently assigned this label at a higher rate than white people. Rather, users felt that the app should \emph{never} assign such a harmful label to images of Black people. By contrast, comparison-based auditing applies to cases where, in order to establish that a harmful bias truly exists and is worth addressing, it is necessary to compare multiple instances and demonstrate the existence of statistical patterns. For example, when Sweeney \cite{sweeney2013discrimination} found that a web search for her own name yielded arrest record ads, she wanted to understand whether systematic discrimination was present. To do so, it was necessary to search for many Black sounding names and statistically compare the results against those for other names \cite{sweeney2013discrimination}. To support such forms of comparative auditing, designers might provide interfaces and algorithmic guidance to empower everyday auditors, who typically lack knowledge of statistics or causal modeling, in testing statistical or causal hypotheses.

\subsection{Design for Organizational Guidance}
In our cases, we also observed fluid divisions of labor among users across the lifetime of an everyday algorithm audit. For example, in the Twitter cropping case, some users largely focused on tweeting images to test how the image cropping algorithm would behave for each. Meanwhile, other users contributed more in retweeting and raising awareness about observed issues, or in forming hypotheses about observed algorithmic behaviors. 

Designers could help guide the division of labor through interventions that encourage everyday auditors to take on particular roles during an auditing process, including roles that are expected to be important and might not arise organically. For example, across all of our cases, a major challenge was a lack of effective synthesis mechanisms to help users keep track of each other’s findings and hypotheses and build upon each other’s contributions so far (cf. \cite{cranshaw2011polymath}). “Synthesizer” may be a valuable role for some everyday auditors to take on, in order to help the collective make more cumulative progress. More broadly, we could imagine organizing everyday auditors along a number of major roles, such as (1) \emph{Initiators}, who focusing on identifying problematic algorithmic behaviors; (2) \emph{Generalizers}, who examinine instances flagged by initiators, and use these to inform hypotheses about the scope of the issue and possible underlying causes; (3) \emph{Amplifiers}, who help share and broadcast what has been discovered to others, raising awareness of findings, hypotheses, and relevant discussions so far; and (4) \emph{Synthesizers}, who transform the outputs of this iterative auditing process into summaries for others to build on. 

\subsection{Design for Incentivization} 
Finally, while community, expert, algorithmic, and organizational guidance may help to improve ongoing auditing processes, another major design opportunity is to explore mechanisms that incentivize everyday auditors to have more active and sustained participation. For example, one potential design direction is to offer an “algorithmic bias bounty program,” drawing inspiration from “bug bounty” programs in cybersecurity. The core idea of security’s “bug bounty” program is to incentivize hackers to share vulnerabilities with legitimate organizations for monetary and reputational rewards, as an alternative to selling or exploiting those vulnerabilities \cite{ozment2004bug, schechter2002buy}. Similarly, in the context of everyday algorithm audits, we could encourage users to directly report detected biases to organizations and platforms, offering social incentives (e.g., reputations and points) and monetary incentives accordingly. A potential risk is that, if not designed carefully, the provision of extrinsic rewards for algorithm auditing may reduce useful spontaneity or may inadvertently diminish users’ existing intrinsic motivations to engage in everyday auditing behaviors.

Indeed, while this paper was under review, Twitter announced its own ``bias bounty'' program, inspired by cybersecurity ``bug bounty'' programs and by its own image cropping cases. This program aimed at incentivizing people to detect, test, and report potential biases \cite{Twitter2021}. However, such ``bias bounty'' programs also heavily rely on participants with relevant technical expertise to detect biases, similar to bug bounty programs in cybersecurity, which rely on experienced hackers to identify bugs. We see design opportunities here to develop systems that can combine strengths of both technical experts and everyday users. For example, an audit could begin by inviting a wide and open range of everyday users to participate in surfacing harmful algorithmic behaviors. Initial hypotheses and observations from everyday users could then serve to inform further, systematic investigations by machine learning experts and analysts. The initial, open phase of auditing may uncover issues that a narrower group of technical experts might otherwise have missed. Indeed, in some ways, this is how the Twitter ``bias bounty'' program proceeded, given that participants in the program were directly informed by observations and hypotheses that everyday users had developed previously.

In addition to incentivizing auditors, there is also a need to incentivize developers and platforms both to implement these measures (e.g., supporting user-led audits on embedded discussion forums) and to remediate issues that users uncover. Holstein et al. (2019) found that commercial product teams, fearful of negative public attention, are often motivated to uncover harmful algorithmic behaviors as early in the process as possible. Acknowledging the limitations of existing auditing approaches, several industry practitioners interviewed in this research expressed interest in implementing mechanisms for user-led auditing \cite{holstein2019improving}. Some product teams were motivated enough to experiment with developing their own collective auditing tools and processes. While these findings highlight that some product teams and companies are already motivated to implement such measures, others may be hesitant to encourage users to scrutinize their systems. Prior work has demonstrated that, even where motivation is lacking, public awareness raising around harmful behaviors in commercial AI systems can motivate target companies to prioritize remediation \cite{raji2019actionable}. Similarly, recent work from Vincent et al. \cite{vincent2021data} discusses potential ways for the public to withhold their data contributions as leverage, reducing the effectiveness of specific data-driven technologies in order to motivate companies to address users’ concerns. Finally, given the uneven power dynamics between users and platforms, structural interventions are also needed (e.g., policy and legal interventions) to ensure external accountability.

\section{Discussion}
We have taken initial steps to theorize and explore an under-studied phenomenon in which everyday users detect, interpret, question, and bring attention to problematic machine behaviors in their day-to-day interactions with algorithms. We argue that such “everyday algorithm auditing” is especially powerful in detecting harmful behaviors that are challenging for existing auditing approaches discussed in the academic literature. In this section, we discuss the unique power and contributions of everyday algorithm auditing and raise a few open questions on how to better support it for future research. 

\subsection{The Power of Everyday Algorithm Auditing}
Past research has highlighted the limitations and blindspots of existing auditing approaches and speculated that everyday users might be able to overcome these limitations (e.g., \cite{eslami2019user,holstein2019improving}). In this paper, we take a step further by studying how such user-led audits unfold in the real world and by exploring how we might better support these processes. Through detailed case studies, we have shown that everyday users are able to detect harmful algorithmic behaviors that are challenging for traditional auditing approaches and that users can exercise agency and autonomy in directing their own auditing process. Below we highlight a number of unique contributions of everyday algorithm auditing and how it can help overcome some of the limitations presented in existing approaches. 

First, everyday algorithm auditing leverages the \emph{lived experiences} of everyday users. Past literature suggests that existing approaches often fail to involve auditors with relevant cultural backgrounds and lived experience that are critical to detect sensitive harms \cite{young2019toward}. Everyday algorithm auditing, with contextually situated users, is especially powerful to appropriately identify the problems at hand. Although a crowdsourced/collaborative auditing approach \cite{sandvig2014auditing} also invites users’ participation, it often relies on crowdworkers, who do not necessarily have lived experiences with the algorithms being audited, do not communicate and work together toward a common goal, and therefore might not be sensitive to the related harms. 

Second, everyday algorithm auditing harnesses the \emph{situated knowledge} of everyday users. As past literature suggests, many harmful machine behaviors are challenging to detect outside of situated contexts of use \cite{cramer2018assessing,friedman1996bias,holstein2019improving,madaio2020co,seaver2017algorithms}. Through their day-to-day interactions with an algorithmic system, everyday users are particularly well positioned to detect these types of behaviors that emerge in real-world contexts of use, in the presence of complex social dynamics, and in the changing norms and practices of using algorithmic systems over time. 

Third, in an everyday algorithm audit, users are able to form \emph{counterpublics} via different social media channels or community forums and participate in their own collective sensemaking and consensus building. This is especially powerful since only in such collective attempts, will users be able to build on each others’ contributions, test each others’ hypotheses, and support each other in different forms (e.g., helping publicize their efforts). In contrast, in a crowdsourced/collaborative audit \cite{sandvig2014auditing}, users are often working on individualized tasks that have been assigned to them, without participating in discussion. 

Finally, in an everyday algorithm audit, users have \emph{control} over the auditing process. This differs from a crowdsourced/collaborative audit \cite{sandvig2014auditing}, in which users are hired via crowdsourcing platforms to work on decomposed subtasks determined entirely by outside parties; in an everyday algorithm auditing, users\thinspace—\thinspace often collectively\thinspace—\thinspace decide their own course of action. Their autonomy and agency help steer the audit in the directions that are most useful and meaningful for them, which are often overlooked by existing approaches. 

\subsection {Appropriate and Timely Interventions} 
We have discussed five types of design interventions we might offer to support everyday algorithm auditing: community guidance, algorithmic guidance, expert guidance, organizational guidance, and incentivization. We have also discussed some potential trade-offs that we need to consider when offering these types of interventions. Below we raise open questions regarding the appropriate timing and proper degree of intervention for future research. 

\subsubsection{When to intervene and when to stop}
We have discussed that interventions can happen at any stage of an everyday algorithm audit and that different stages might benefit from different types of interventions. For example, we might anticipate that having expert guidance during the initiation stage (e.g., as in the case of ImageNet Roulette) and the hypothesizing and testing stage (e.g., in the Twitter racial cropping case) might be especially helpful. On the other hand, community guidance might be more useful at the awareness raising stage (e.g., as in the YouTube LGBTQ+ demonetization case) and the hypothesizing and testing stage (e.g., in the Yelp and Twitter cases). However, currently we have very limited understanding of the appropriate timing for intervention. For example, how can we maintain the organic nature of everyday algorithm auditing, especially at the initiation stage, but also prompt a group of people to start auditing first, which might inevitably remove some of the organic nature? 

\subsubsection{How much intervention}
We have also discussed different forms of intervention and some hypothesized trade-offs. For example, introducing expert guidance might create unequal power dynamics in community discussions and risk diminishing users’ sense of community and autonomy. Hosting community discussions within the same platforms that are under audit may lead to concerns about platform-driven censorship or suppression. Currently we are only at the starting point to understand and explore the appropriate degrees of these interventions. How much intervention is too much? What are the trade-offs in moving from more organic to more organized at different phases of an everyday audit? These questions need further research and investigation; the exploration of everyday algorithm auditing is in its infancy, and our work presents a first step .

\section{Conclusions}
In this paper, we have drawn on real-world case studies and prior theories of everyday resistance and counterpublics to understand how everyday users\thinspace—\thinspace either individually or collectively\thinspace—\thinspace detect, understand, and interrogate problematic machine behaviors in their day-to-day interactions with algorithmic systems. By comparing the lifetime and dynamics of several cases of everyday algorithm auditing, we have proposed a process-oriented view of everyday algorithm audits, involving initiation, awareness raising, testing and hypothesizing, and remediation. We have drawn lessons from these cases for future research and design around everyday algorithm auditing, outlining five broad categories of potential design interventions to support everyday audits\thinspace—\thinspace (a) community guidance, (b) expert guidance, (c) algorithmic guidance, (d) organizational guidance, and (e) incentivization\thinspace—\thinspace along with potential trade-offs.

As the first work conceptualizing and studying everyday algorithm auditing, this research is highly exploratory in nature. The lifetime and dynamics presented as part of an everyday audit are intended to be representations of what we see at this stage of exploration, capturing both current practices we have observed and possibilities we see for future practice. As such, these represent a set of untested assumptions about everyday algorithm audits that could change over time as more empirical work is conducted in this space. That is, neither the cases presented in this paper nor the approaches we used to analyze them are comprehensive: they outline an observed phenomenon and serve as illustrative examples to better understand how everyday audits work, what paths everyday audits could take, and what impacts everyday audits could have. Future research should conduct in-depth follow-up studies with users to investigate the practices and strategies they use to find and make sense of harmful algorithmic behaviors. Another fruitful, complementary area for future research includes an exploration of the design opportunities discussed above. For example, how can we design platforms that support users in conducting more effective everyday algorithm audits, both individually and collectively? 

In sum, we view this work as an initial step toward bridging the gap between algorithm auditing approaches in academia/industry and everyday auditing behaviors that emerge in day-to-day use of algorithmic systems. It is our hope that this work will help to inform future research on everyday algorithm auditing, as well as the design of future platforms and tools that empower people to meaningfully audit the algorithmic systems that impact their lives every day.

\begin{acks}
This work was supported by the National Science Foundation (NSF) program on Fairness in AI in collaboration with Amazon under Award No. IIS-2040942, an Amazon Research Award and a Cisco Research Award. We thank Jason Hong, Adam Perer, Nihar Shah and anonymous reviewers for offering helpful comments. 
\end{acks}

\bibliographystyle{ACM-Reference-Format}
\bibliography{auditing-minor}


\begin{thebibliography}{101}


\ifx \showCODEN    \undefined \def \showCODEN     #1{\unskip}     \fi
\ifx \showDOI      \undefined \def \showDOI       #1{#1}\fi
\ifx \showISBNx    \undefined \def \showISBNx     #1{\unskip}     \fi
\ifx \showISBNxiii \undefined \def \showISBNxiii  #1{\unskip}     \fi
\ifx \showISSN     \undefined \def \showISSN      #1{\unskip}     \fi
\ifx \showLCCN     \undefined \def \showLCCN      #1{\unskip}     \fi
\ifx \shownote     \undefined \def \shownote      #1{#1}          \fi
\ifx \showarticletitle \undefined \def \showarticletitle #1{#1}   \fi
\ifx \showURL      \undefined \def \showURL       {\relax}        \fi
\providecommand\bibfield[2]{#2}
\providecommand\bibinfo[2]{#2}
\providecommand\natexlab[1]{#1}
\providecommand\showeprint[2][]{arXiv:#2}

\bibitem[\protect\citeauthoryear{??}{bbc}{2020}]%
        {bbc2020twitter}
 \bibinfo{year}{2020}\natexlab{}.
\newblock \showarticletitle{Twitter investigates racial bias in image
  previews}.
\newblock \bibinfo{journal}{\emph{BBC News}} (\bibinfo{date}{Sep}
  \bibinfo{year}{2020}).
\newblock
\urldef\tempurl%
\url{https://www.bbc.com/news/technology-54234822}
\showURL{%
\tempurl}


\bibitem[\protect\citeauthoryear{??}{AJL}{2021}]%
        {AJLmission}
 \bibinfo{year}{2021}\natexlab{}.
\newblock \bibinfo{title}{The Algorithmic Justice League: Mission, Team and
  Story}.
\newblock
\newblock
\urldef\tempurl%
\url{https://www.ajl.org/about}
\showURL{%
\tempurl}


\bibitem[\protect\citeauthoryear{Agrawal and Davis}{Agrawal and Davis}{2020}]%
        {Twitter2020}
\bibfield{author}{\bibinfo{person}{Parag Agrawal} {and}
  \bibinfo{person}{Dantley Davis}.} \bibinfo{year}{2020}\natexlab{}.
\newblock \showarticletitle{Transparency around image cropping and changes to
  come}.
\newblock \bibinfo{journal}{\emph{Twitter}} (\bibinfo{year}{2020}).
\newblock
\urldef\tempurl%
\url{https://blog.twitter.com/en_us/topics/product/2020/transparency-image-cropping.html.}
\showURL{%
\tempurl}


\bibitem[\protect\citeauthoryear{Anderson and Magruder}{Anderson and
  Magruder}{2012}]%
        {anderson2012learning}
\bibfield{author}{\bibinfo{person}{Michael Anderson} {and}
  \bibinfo{person}{Jeremy Magruder}.} \bibinfo{year}{2012}\natexlab{}.
\newblock \showarticletitle{Learning from the crowd: Regression discontinuity
  estimates of the effects of an online review database}.
\newblock \bibinfo{journal}{\emph{The Economic Journal}} \bibinfo{volume}{122},
  \bibinfo{number}{563} (\bibinfo{year}{2012}), \bibinfo{pages}{957--989}.
\newblock


\bibitem[\protect\citeauthoryear{Asplund, Eslami, Sundaram, Sandvig, and
  Karahalios}{Asplund et~al\mbox{.}}{2020}]%
        {asplund2020auditing}
\bibfield{author}{\bibinfo{person}{Joshua Asplund}, \bibinfo{person}{Motahhare
  Eslami}, \bibinfo{person}{Hari Sundaram}, \bibinfo{person}{Christian
  Sandvig}, {and} \bibinfo{person}{Karrie Karahalios}.}
  \bibinfo{year}{2020}\natexlab{}.
\newblock \showarticletitle{Auditing race and gender discrimination in online
  housing markets}. In \bibinfo{booktitle}{\emph{Proceedings of the
  International AAAI Conference on Web and Social Media}},
  Vol.~\bibinfo{volume}{14}. \bibinfo{pages}{24--35}.
\newblock


\bibitem[\protect\citeauthoryear{Bansal and Weld}{Bansal and Weld}{2018}]%
        {bansal2018coverage}
\bibfield{author}{\bibinfo{person}{Gagan Bansal} {and} \bibinfo{person}{Daniel
  Weld}.} \bibinfo{year}{2018}\natexlab{}.
\newblock \showarticletitle{A coverage-based utility model for identifying
  unknown unknowns}. In \bibinfo{booktitle}{\emph{Proceedings of the AAAI
  Conference on Artificial Intelligence}}, Vol.~\bibinfo{volume}{32}.
\newblock


\bibitem[\protect\citeauthoryear{Bishop}{Bishop}{2015}]%
        {bishop2015amazon}
\bibfield{author}{\bibinfo{person}{Todd Bishop}.}
  \bibinfo{year}{2015}\natexlab{}.
\newblock \showarticletitle{Amazon changes its key formula for calculating
  product ratings and displaying reviews}.
\newblock \bibinfo{journal}{\emph{GeekWire}} (\bibinfo{date}{June}
  \bibinfo{year}{2015}).
\newblock
\urldef\tempurl%
\url{https://www.geekwire.com/2015/amazon-changes-its-influential-formula-for-calculating-product-ratings}
\showURL{%
\tempurl}


\bibitem[\protect\citeauthoryear{Blodgett, Barocas, Daum{\'e}~III, and
  Wallach}{Blodgett et~al\mbox{.}}{2020}]%
        {blodgett2020language}
\bibfield{author}{\bibinfo{person}{Su~Lin Blodgett}, \bibinfo{person}{Solon
  Barocas}, \bibinfo{person}{Hal Daum{\'e}~III}, {and} \bibinfo{person}{Hanna
  Wallach}.} \bibinfo{year}{2020}\natexlab{}.
\newblock \showarticletitle{Language (technology) is power: A critical survey
  of" bias" in nlp}.
\newblock \bibinfo{journal}{\emph{arXiv preprint arXiv:2005.14050}}
  (\bibinfo{year}{2020}).
\newblock


\bibitem[\protect\citeauthoryear{Brian}{Brian}{2019}]%
        {brian2019booking}
\bibfield{author}{\bibinfo{person}{Rucksack Brian}.}
  \bibinfo{year}{2019}\natexlab{}.
\newblock \bibinfo{title}{Major changes to review scoring method on Booking
  com}.
\newblock
\newblock
\urldef\tempurl%
\url{https://hostelmanagement.com/forums/major-changes-review-scoring-method-booking-com.html}
\showURL{%
\tempurl}


\bibitem[\protect\citeauthoryear{Brooks}{Brooks}{2020}]%
        {brooks2020twitter}
\bibfield{author}{\bibinfo{person}{Khristopher~J Brooks}.}
  \bibinfo{year}{2020}\natexlab{}.
\newblock \showarticletitle{Twitter users say the platform crops out Black
  faces}.
\newblock \bibinfo{journal}{\emph{CBS News}} (\bibinfo{date}{Sep}
  \bibinfo{year}{2020}).
\newblock
\urldef\tempurl%
\url{https://www.cbsnews.com/news/twitter-image-cropping-algorithm-racial-profiling/}
\showURL{%
\tempurl}


\bibitem[\protect\citeauthoryear{Buolamwini}{Buolamwini}{2016}]%
        {buolamwini2016m}
\bibfield{author}{\bibinfo{person}{Joy Buolamwini}.}
  \bibinfo{year}{2016}\natexlab{}.
\newblock \showarticletitle{How I’m fighting bias in algorithms}.
\newblock \bibinfo{journal}{\emph{November, TEDx Beacon Street)[Video File]}}
  (\bibinfo{year}{2016}).
\newblock


\bibitem[\protect\citeauthoryear{Buolamwini and Gebru}{Buolamwini and
  Gebru}{2018}]%
        {buolamwini2018gender}
\bibfield{author}{\bibinfo{person}{Joy Buolamwini} {and}
  \bibinfo{person}{Timnit Gebru}.} \bibinfo{year}{2018}\natexlab{}.
\newblock \showarticletitle{Gender shades: Intersectional accuracy disparities
  in commercial gender classification}. In
  \bibinfo{booktitle}{\emph{Proceedings of the 1st Conference on Fairness,
  Accountability and Transparency PMLR}}. \bibinfo{pages}{77--91}.
\newblock


\bibitem[\protect\citeauthoryear{Chan, Dang, and Dow}{Chan
  et~al\mbox{.}}{2016}]%
        {chan2016improving}
\bibfield{author}{\bibinfo{person}{Joel Chan}, \bibinfo{person}{Steven Dang},
  {and} \bibinfo{person}{Steven~P Dow}.} \bibinfo{year}{2016}\natexlab{}.
\newblock \showarticletitle{Improving crowd innovation with expert
  facilitation}. In \bibinfo{booktitle}{\emph{Proceedings of the 19th ACM
  Conference on Computer-Supported Cooperative Work \& Social Computing}}.
  \bibinfo{pages}{1223--1235}.
\newblock


\bibitem[\protect\citeauthoryear{Chen, Ma, Hann{\'a}k, and Wilson}{Chen
  et~al\mbox{.}}{2018}]%
        {chen2018investigating}
\bibfield{author}{\bibinfo{person}{Le Chen}, \bibinfo{person}{Ruijun Ma},
  \bibinfo{person}{Anik{\'o} Hann{\'a}k}, {and} \bibinfo{person}{Christo
  Wilson}.} \bibinfo{year}{2018}\natexlab{}.
\newblock \showarticletitle{Investigating the impact of gender on rank in
  resume search engines}. In \bibinfo{booktitle}{\emph{Proceedings of the 2018
  CHI Conference on Human Factors in Computing Systems}}.
  \bibinfo{pages}{1--14}.
\newblock


\bibitem[\protect\citeauthoryear{Chen, Cao, Xu, Cheng, Wang, and Li}{Chen
  et~al\mbox{.}}{2020}]%
        {chen2020understanding}
\bibfield{author}{\bibinfo{person}{Zhilong Chen}, \bibinfo{person}{Hancheng
  Cao}, \bibinfo{person}{Fengli Xu}, \bibinfo{person}{Mengjie Cheng},
  \bibinfo{person}{Tao Wang}, {and} \bibinfo{person}{Yong Li}.}
  \bibinfo{year}{2020}\natexlab{}.
\newblock \showarticletitle{Understanding the Role of Intermediaries in Online
  Social E-commerce: An Exploratory Study of Beidian}.
\newblock \bibinfo{journal}{\emph{Proceedings of the ACM on Human-Computer
  Interaction}} \bibinfo{volume}{4}, \bibinfo{number}{CSCW2}
  (\bibinfo{year}{2020}), \bibinfo{pages}{1--24}.
\newblock


\bibitem[\protect\citeauthoryear{Chowdhury and Williams}{Chowdhury and
  Williams}{2021}]%
        {Twitter2021}
\bibfield{author}{\bibinfo{person}{Rumman Chowdhury} {and}
  \bibinfo{person}{Jutta Williams}.} \bibinfo{year}{2021}\natexlab{}.
\newblock \showarticletitle{Introducing Twitter’s first algorithmic bias
  bounty challenge}.
\newblock \bibinfo{journal}{\emph{Twitter}} (\bibinfo{year}{2021}).
\newblock
\urldef\tempurl%
\url{https://blog.twitter.com/engineering/en_us/topics/insights/2021/algorithmic-bias-bounty-challenge}
\showURL{%
\tempurl}


\bibitem[\protect\citeauthoryear{Cramer, Garcia-Gathright, Springer, and
  Reddy}{Cramer et~al\mbox{.}}{2018}]%
        {cramer2018assessing}
\bibfield{author}{\bibinfo{person}{Henriette Cramer}, \bibinfo{person}{Jean
  Garcia-Gathright}, \bibinfo{person}{Aaron Springer}, {and}
  \bibinfo{person}{Sravana Reddy}.} \bibinfo{year}{2018}\natexlab{}.
\newblock \showarticletitle{Assessing and addressing algorithmic bias in
  practice}.
\newblock \bibinfo{journal}{\emph{Interactions}} \bibinfo{volume}{25},
  \bibinfo{number}{6} (\bibinfo{year}{2018}), \bibinfo{pages}{58--63}.
\newblock


\bibitem[\protect\citeauthoryear{Cranshaw and Kittur}{Cranshaw and
  Kittur}{2011}]%
        {cranshaw2011polymath}
\bibfield{author}{\bibinfo{person}{Justin Cranshaw} {and}
  \bibinfo{person}{Aniket Kittur}.} \bibinfo{year}{2011}\natexlab{}.
\newblock \showarticletitle{The polymath project: lessons from a successful
  online collaboration in mathematics}. In
  \bibinfo{booktitle}{\emph{Proceedings of the 2011 CHI conference on Human
  Factors in Computing Systems}}. \bibinfo{pages}{1865--1874}.
\newblock


\bibitem[\protect\citeauthoryear{Crawford and Paglen}{Crawford and
  Paglen}{2021}]%
        {crawford2019excavating}
\bibfield{author}{\bibinfo{person}{Kate Crawford} {and} \bibinfo{person}{Trevor
  Paglen}.} \bibinfo{year}{2021}\natexlab{}.
\newblock \showarticletitle{Excavating AI: The politics of images in machine
  learning training sets}.
\newblock \bibinfo{journal}{\emph{AI \& Society}} (\bibinfo{year}{2021}),
  \bibinfo{pages}{1--12}.
\newblock


\bibitem[\protect\citeauthoryear{Dastin}{Dastin}{2018}]%
        {dastin2018amazon}
\bibfield{author}{\bibinfo{person}{Jeffrey Dastin}.}
  \bibinfo{year}{2018}\natexlab{}.
\newblock \showarticletitle{Amazon scraps secret AI recruiting tool that showed
  bias against women}.
\newblock \bibinfo{journal}{\emph{Reuters}} (\bibinfo{date}{Oct.}
  \bibinfo{year}{2018}).
\newblock
\urldef\tempurl%
\url{https://www.reuters.com/article/us-amazon-com-jobs-automation-insight/amazon-scraps-secret-ai-recruiting-tool-that-showed-bias-against-women-idUSKCN1MK08G}
\showURL{%
\tempurl}


\bibitem[\protect\citeauthoryear{Dawkins}{Dawkins}{2016}]%
        {dawkins2016selfish}
\bibfield{author}{\bibinfo{person}{Richard Dawkins}.}
  \bibinfo{year}{2016}\natexlab{}.
\newblock \bibinfo{booktitle}{\emph{The selfish gene}}.
\newblock \bibinfo{publisher}{Oxford University Press}.
\newblock


\bibitem[\protect\citeauthoryear{De~Certeau}{De~Certeau}{1984}]%
        {de1984practice}
\bibfield{author}{\bibinfo{person}{Michel De~Certeau}.}
  \bibinfo{year}{1984}\natexlab{}.
\newblock \bibinfo{booktitle}{\emph{The practice of everyday life, trans Steven
  Rendall}}.
\newblock \bibinfo{publisher}{Berkeley: University of California Press}.
\newblock


\bibitem[\protect\citeauthoryear{DeVito, Birnholtz, Hancock, French, and
  Liu}{DeVito et~al\mbox{.}}{2018}]%
        {devito2018people}
\bibfield{author}{\bibinfo{person}{Michael~A DeVito}, \bibinfo{person}{Jeremy
  Birnholtz}, \bibinfo{person}{Jeffery~T Hancock}, \bibinfo{person}{Megan
  French}, {and} \bibinfo{person}{Sunny Liu}.} \bibinfo{year}{2018}\natexlab{}.
\newblock \showarticletitle{How people form folk theories of social media feeds
  and what it means for how we study self-presentation}. In
  \bibinfo{booktitle}{\emph{Proceedings of the 2018 CHI Conference on Human
  Factors in Computing Systems}}. \bibinfo{pages}{1--12}.
\newblock


\bibitem[\protect\citeauthoryear{DeVito, Gergle, and Birnholtz}{DeVito
  et~al\mbox{.}}{2017}]%
        {devito2017algorithms}
\bibfield{author}{\bibinfo{person}{Michael~A DeVito}, \bibinfo{person}{Darren
  Gergle}, {and} \bibinfo{person}{Jeremy Birnholtz}.}
  \bibinfo{year}{2017}\natexlab{}.
\newblock \showarticletitle{``Algorithms ruin everything'' \# RIPTwitter, Folk
  Theories, and Resistance to Algorithmic Change in Social Media}. In
  \bibinfo{booktitle}{\emph{Proceedings of the 2017 CHI Conference on Human
  Factors in Computing Systems}}. \bibinfo{pages}{3163--3174}.
\newblock


\bibitem[\protect\citeauthoryear{Diakopoulos}{Diakopoulos}{2014}]%
        {diakopoulos2014algorithmic}
\bibfield{author}{\bibinfo{person}{Nicholas Diakopoulos}.}
  \bibinfo{year}{2014}\natexlab{}.
\newblock \showarticletitle{Algorithmic accountability reporting: On the
  investigation of black boxes}.
\newblock \bibinfo{journal}{\emph{The Tow Center for Digital Journalism}}
  (\bibinfo{year}{2014}).
\newblock


\bibitem[\protect\citeauthoryear{Dow, Kulkarni, Klemmer, and Hartmann}{Dow
  et~al\mbox{.}}{2012}]%
        {dow2012shepherding}
\bibfield{author}{\bibinfo{person}{Steven Dow}, \bibinfo{person}{Anand
  Kulkarni}, \bibinfo{person}{Scott Klemmer}, {and} \bibinfo{person}{Bj{\"o}rn
  Hartmann}.} \bibinfo{year}{2012}\natexlab{}.
\newblock \showarticletitle{Shepherding the crowd yields better work}. In
  \bibinfo{booktitle}{\emph{Proceedings of the ACM 2012 Conference on Computer
  Supported Cooperative Work}}. \bibinfo{pages}{1013--1022}.
\newblock


\bibitem[\protect\citeauthoryear{Eslami, Karahalios, Sandvig, Vaccaro, Rickman,
  Hamilton, and Kirlik}{Eslami et~al\mbox{.}}{2016}]%
        {eslami2016first}
\bibfield{author}{\bibinfo{person}{Motahhare Eslami}, \bibinfo{person}{Karrie
  Karahalios}, \bibinfo{person}{Christian Sandvig}, \bibinfo{person}{Kristen
  Vaccaro}, \bibinfo{person}{Aimee Rickman}, \bibinfo{person}{Kevin Hamilton},
  {and} \bibinfo{person}{Alex Kirlik}.} \bibinfo{year}{2016}\natexlab{}.
\newblock \showarticletitle{First I ``like'' it, then I hide it: Folk theories
  of social feeds}. In \bibinfo{booktitle}{\emph{Proceedings of the 2016 CHI
  Conference on Human Factors in Computing Systems}}.
  \bibinfo{pages}{2371--2382}.
\newblock


\bibitem[\protect\citeauthoryear{Eslami, Vaccaro, Karahalios, and
  Hamilton}{Eslami et~al\mbox{.}}{2017}]%
        {eslami2017careful}
\bibfield{author}{\bibinfo{person}{Motahhare Eslami}, \bibinfo{person}{Kristen
  Vaccaro}, \bibinfo{person}{Karrie Karahalios}, {and} \bibinfo{person}{Kevin
  Hamilton}.} \bibinfo{year}{2017}\natexlab{}.
\newblock \showarticletitle{“Be careful; things can be worse than they
  appear”: Understanding Biased Algorithms and Users’ Behavior around Them
  in Rating Platforms}. In \bibinfo{booktitle}{\emph{Proceedings of the
  International AAAI Conference on Web and Social Media}},
  Vol.~\bibinfo{volume}{11}.
\newblock


\bibitem[\protect\citeauthoryear{Eslami, Vaccaro, Lee, Elazari Bar~On, Gilbert,
  and Karahalios}{Eslami et~al\mbox{.}}{2019}]%
        {eslami2019user}
\bibfield{author}{\bibinfo{person}{Motahhare Eslami}, \bibinfo{person}{Kristen
  Vaccaro}, \bibinfo{person}{Min~Kyung Lee}, \bibinfo{person}{Amit Elazari
  Bar~On}, \bibinfo{person}{Eric Gilbert}, {and} \bibinfo{person}{Karrie
  Karahalios}.} \bibinfo{year}{2019}\natexlab{}.
\newblock \showarticletitle{User attitudes towards algorithmic opacity and
  transparency in online reviewing platforms}. In
  \bibinfo{booktitle}{\emph{Proceedings of the 2019 CHI Conference on Human
  Factors in Computing Systems}}. \bibinfo{pages}{1--14}.
\newblock


\bibitem[\protect\citeauthoryear{Fisher, Counts, and Kittur}{Fisher
  et~al\mbox{.}}{2012}]%
        {fisher2012distributed}
\bibfield{author}{\bibinfo{person}{Kristie Fisher}, \bibinfo{person}{Scott
  Counts}, {and} \bibinfo{person}{Aniket Kittur}.}
  \bibinfo{year}{2012}\natexlab{}.
\newblock \showarticletitle{Distributed sensemaking: Improving sensemaking by
  leveraging the efforts of previous users}. In
  \bibinfo{booktitle}{\emph{Proceedings of the 2012 CHI Conference on Human
  Factors in Computing Systems}}. \bibinfo{pages}{247--256}.
\newblock


\bibitem[\protect\citeauthoryear{Fowler}{Fowler}{2011}]%
        {fowler2011judge}
\bibfield{author}{\bibinfo{person}{Geoffrey~A. Fowler}.}
  \bibinfo{year}{2011}\natexlab{}.
\newblock \showarticletitle{Judge dismisses suit against Yelp}.
\newblock \bibinfo{journal}{\emph{The Wall Street Journal}}
  (\bibinfo{date}{Oct.} \bibinfo{year}{2011}).
\newblock
\urldef\tempurl%
\url{https://www.wsj.com/articles/SB10001424052970204505304577002170423750412}
\showURL{%
\tempurl}


\bibitem[\protect\citeauthoryear{Fraser}{Fraser}{1990}]%
        {fraser1990rethinking}
\bibfield{author}{\bibinfo{person}{Nancy Fraser}.}
  \bibinfo{year}{1990}\natexlab{}.
\newblock \showarticletitle{Rethinking the public sphere: A contribution to the
  critique of actually existing democracy}.
\newblock \bibinfo{journal}{\emph{Social text}} \bibinfo{number}{25/26}
  (\bibinfo{year}{1990}), \bibinfo{pages}{56--80}.
\newblock


\bibitem[\protect\citeauthoryear{Friedman and Nissenbaum}{Friedman and
  Nissenbaum}{1996}]%
        {friedman1996bias}
\bibfield{author}{\bibinfo{person}{Batya Friedman} {and} \bibinfo{person}{Helen
  Nissenbaum}.} \bibinfo{year}{1996}\natexlab{}.
\newblock \showarticletitle{Bias in computer systems}.
\newblock \bibinfo{journal}{\emph{ACM Transactions on Information Systems
  (TOIS)}} \bibinfo{volume}{14}, \bibinfo{number}{3} (\bibinfo{year}{1996}),
  \bibinfo{pages}{330--347}.
\newblock


\bibitem[\protect\citeauthoryear{Fung}{Fung}{2015}]%
        {Fung2015map}
\bibfield{author}{\bibinfo{person}{Brian Fung}.}
  \bibinfo{year}{2015}\natexlab{}.
\newblock \showarticletitle{The Internet has unearthed more racist Google Maps
  results}.
\newblock \bibinfo{journal}{\emph{The Washington Post}} (\bibinfo{date}{May}
  \bibinfo{year}{2015}).
\newblock
\urldef\tempurl%
\url{https://www.washingtonpost.com/news/the-switch/wp/2015/05/21/the-internet-has-unearthed-more-racist-google-maps-results}
\showURL{%
\tempurl}


\bibitem[\protect\citeauthoryear{Geiger}{Geiger}{2016}]%
        {geiger2016bot}
\bibfield{author}{\bibinfo{person}{R~Stuart Geiger}.}
  \bibinfo{year}{2016}\natexlab{}.
\newblock \showarticletitle{Bot-based collective blocklists in Twitter: the
  counterpublic moderation of harassment in a networked public space}.
\newblock \bibinfo{journal}{\emph{Information, Communication \& Society}}
  \bibinfo{volume}{19}, \bibinfo{number}{6} (\bibinfo{year}{2016}),
  \bibinfo{pages}{787--803}.
\newblock


\bibitem[\protect\citeauthoryear{Geiger and Halfaker}{Geiger and
  Halfaker}{2013}]%
        {geiger2013using}
\bibfield{author}{\bibinfo{person}{R~Stuart Geiger} {and}
  \bibinfo{person}{Aaron Halfaker}.} \bibinfo{year}{2013}\natexlab{}.
\newblock \showarticletitle{Using edit sessions to measure participation in
  Wikipedia}. In \bibinfo{booktitle}{\emph{Proceedings of the 2013 Conference
  on Computer Supported Cooperative Work}}. \bibinfo{pages}{861--870}.
\newblock


\bibitem[\protect\citeauthoryear{Gillespie}{Gillespie}{2014}]%
        {gillespie2014relevance}
\bibfield{author}{\bibinfo{person}{Tarleton Gillespie}.}
  \bibinfo{year}{2014}\natexlab{}.
\newblock \showarticletitle{The relevance of algorithms}.
\newblock \bibinfo{journal}{\emph{Media technologies: Essays on communication,
  materiality, and society}} \bibinfo{volume}{167}, \bibinfo{number}{2014}
  (\bibinfo{year}{2014}), \bibinfo{pages}{167}.
\newblock


\bibitem[\protect\citeauthoryear{Goldman}{Goldman}{2014}]%
        {goldman2014yelp}
\bibfield{author}{\bibinfo{person}{Eric Goldman}.}
  \bibinfo{year}{2014}\natexlab{}.
\newblock \showarticletitle{Court says Yelp doesn't extort businesses}.
\newblock \bibinfo{journal}{\emph{Forbes}} (\bibinfo{date}{Sept.}
  \bibinfo{year}{2014}).
\newblock
\urldef\tempurl%
\url{https://www.forbes.com/sites/ericgoldman/2014/09/03/court-says-yelp-doesnt-extort-businesses}
\showURL{%
\tempurl}


\bibitem[\protect\citeauthoryear{Goodman and Flaxman}{Goodman and
  Flaxman}{2016}]%
        {goodman2016eu}
\bibfield{author}{\bibinfo{person}{Bryce Goodman} {and} \bibinfo{person}{Seth
  Flaxman}.} \bibinfo{year}{2016}\natexlab{}.
\newblock \showarticletitle{EU regulations on algorithmic decision-making and a
  “right to explanation”}. In \bibinfo{booktitle}{\emph{ICML Workshop on
  Human Interpretability in Machine Learning}}.
\newblock


\bibitem[\protect\citeauthoryear{Grevet and Gilbert}{Grevet and
  Gilbert}{2015}]%
        {grevet2015piggyback}
\bibfield{author}{\bibinfo{person}{Catherine Grevet} {and}
  \bibinfo{person}{Eric Gilbert}.} \bibinfo{year}{2015}\natexlab{}.
\newblock \showarticletitle{Piggyback prototyping: Using existing, large-scale
  social computing systems to prototype new ones}. In
  \bibinfo{booktitle}{\emph{Proceedings of the 2015 CHI Conference on Human
  Factors in Computing Systems}}. \bibinfo{pages}{4047--4056}.
\newblock


\bibitem[\protect\citeauthoryear{Guadagno, Rempala, Murphy, and Okdie}{Guadagno
  et~al\mbox{.}}{2013}]%
        {guadagno2013what}
\bibfield{author}{\bibinfo{person}{Rosanna~E. Guadagno},
  \bibinfo{person}{Daniel~M. Rempala}, \bibinfo{person}{Shannon Murphy}, {and}
  \bibinfo{person}{Bradley~M. Okdie}.} \bibinfo{year}{2013}\natexlab{}.
\newblock \showarticletitle{What makes a video go viral? An analysis of
  emotional contagion and Internet memes}.
\newblock \bibinfo{journal}{\emph{Computers in Human Behavior}}
  \bibinfo{volume}{29}, \bibinfo{number}{6} (\bibinfo{year}{2013}),
  \bibinfo{pages}{2312--2319}.
\newblock
\showISSN{0747-5632}
\urldef\tempurl%
\url{https://doi.org/10.1016/j.chb.2013.04.016}
\showDOI{\tempurl}


\bibitem[\protect\citeauthoryear{Guynn}{Guynn}{2015}]%
        {guynn2015google}
\bibfield{author}{\bibinfo{person}{Jessica Guynn}.}
  \bibinfo{year}{2015}\natexlab{}.
\newblock \showarticletitle{Google photos labeled black people `gorillas'}.
\newblock \bibinfo{journal}{\emph{USA Today}} (\bibinfo{date}{June}
  \bibinfo{year}{2015}).
\newblock
\urldef\tempurl%
\url{https://www.usatoday.com/story/tech/2015/07/01/google-apologizes-after-photos-identify-black-people-as-gorillas/29567465}
\showURL{%
\tempurl}


\bibitem[\protect\citeauthoryear{Habermas}{Habermas}{1989}]%
        {habermas1991structural}
\bibfield{author}{\bibinfo{person}{J{\"u}rgen Habermas}.}
  \bibinfo{year}{1989}\natexlab{}.
\newblock \bibinfo{booktitle}{\emph{The structural transformation of the public
  sphere: An inquiry into a category of bourgeois society}}.
\newblock \bibinfo{publisher}{MIT press}.
\newblock


\bibitem[\protect\citeauthoryear{Halfaker, Kittur, and Riedl}{Halfaker
  et~al\mbox{.}}{2011}]%
        {halfaker2011don}
\bibfield{author}{\bibinfo{person}{Aaron Halfaker}, \bibinfo{person}{Aniket
  Kittur}, {and} \bibinfo{person}{John Riedl}.}
  \bibinfo{year}{2011}\natexlab{}.
\newblock \showarticletitle{Don't bite the newbies: How reverts affect the
  quantity and quality of Wikipedia work}. In
  \bibinfo{booktitle}{\emph{Proceedings of the 7th International Symposium on
  Wikis and Open Collaboration}}. \bibinfo{pages}{163--172}.
\newblock


\bibitem[\protect\citeauthoryear{Hannak, Soeller, Lazer, Mislove, and
  Wilson}{Hannak et~al\mbox{.}}{2014}]%
        {hannak2014measuring}
\bibfield{author}{\bibinfo{person}{Aniko Hannak}, \bibinfo{person}{Gary
  Soeller}, \bibinfo{person}{David Lazer}, \bibinfo{person}{Alan Mislove},
  {and} \bibinfo{person}{Christo Wilson}.} \bibinfo{year}{2014}\natexlab{}.
\newblock \showarticletitle{Measuring price discrimination and steering on
  e-commerce web sites}. In \bibinfo{booktitle}{\emph{Proceedings of the 2014
  Conference on Internet Measurement Conference}}. \bibinfo{pages}{305--318}.
\newblock


\bibitem[\protect\citeauthoryear{Hern}{Hern}{2020}]%
        {guardian2020}
\bibfield{author}{\bibinfo{person}{Alex Hern}.}
  \bibinfo{year}{2020}\natexlab{}.
\newblock \showarticletitle{Twitter apologises for 'racist' image-cropping
  algorithm}.
\newblock \bibinfo{journal}{\emph{The Guardian}} (\bibinfo{date}{Sept.}
  \bibinfo{year}{2020}).
\newblock
\urldef\tempurl%
\url{https://www.theguardian.com/technology/2020/sep/21/twitter-apologises-for-racist-image-cropping-algorithm}
\showURL{%
\tempurl}


\bibitem[\protect\citeauthoryear{Holstein, Harpstead, Gulotta, and
  Forlizzi}{Holstein et~al\mbox{.}}{2020}]%
        {holstein2020replay}
\bibfield{author}{\bibinfo{person}{Kenneth Holstein}, \bibinfo{person}{Erik
  Harpstead}, \bibinfo{person}{Rebecca Gulotta}, {and} \bibinfo{person}{Jodi
  Forlizzi}.} \bibinfo{year}{2020}\natexlab{}.
\newblock \showarticletitle{Replay enactments: Exploring possible futures
  through historical data}. In \bibinfo{booktitle}{\emph{Proceedings of the
  2020 ACM Designing Interactive Systems Conference}}.
  \bibinfo{pages}{1607--1618}.
\newblock


\bibitem[\protect\citeauthoryear{Holstein, Wortman~Vaughan, Daum{\'e}~III,
  Dudik, and Wallach}{Holstein et~al\mbox{.}}{2019}]%
        {holstein2019improving}
\bibfield{author}{\bibinfo{person}{Kenneth Holstein}, \bibinfo{person}{Jennifer
  Wortman~Vaughan}, \bibinfo{person}{Hal Daum{\'e}~III}, \bibinfo{person}{Miro
  Dudik}, {and} \bibinfo{person}{Hanna Wallach}.}
  \bibinfo{year}{2019}\natexlab{}.
\newblock \showarticletitle{Improving fairness in machine learning systems:
  What do industry practitioners need?}. In
  \bibinfo{booktitle}{\emph{Proceedings of the 2019 CHI Conference on Human
  Factors in Computing Systems}}. \bibinfo{pages}{1--16}.
\newblock


\bibitem[\protect\citeauthoryear{House}{House}{2016}]%
        {house2016big}
\bibfield{author}{\bibinfo{person}{White House}.}
  \bibinfo{year}{2016}\natexlab{}.
\newblock \showarticletitle{Big Data: A Report on Algorithmic Systems}.
\newblock \bibinfo{journal}{\emph{Opportunity, and Civil Rights}}
  (\bibinfo{year}{2016}).
\newblock


\bibitem[\protect\citeauthoryear{Johnson}{Johnson}{2019}]%
        {johnson2019ai}
\bibfield{author}{\bibinfo{person}{Khari Johnson}.}
  \bibinfo{year}{2019}\natexlab{}.
\newblock \showarticletitle{AI predictions for 2019 from Yann LeCun, Hilary
  Mason, Andrew Ng, and Rumman Chowdhury}.
\newblock \bibinfo{journal}{\emph{VentureBeat}} (\bibinfo{date}{Jan}
  \bibinfo{year}{2019}).
\newblock
\urldef\tempurl%
\url{https://venturebeat.com/2019/01/02/ai-predictions-for-2019-from-yann-lecun-hilary-mason-andrew-ng-and-rumman-chowdhury/}
\showURL{%
\tempurl}


\bibitem[\protect\citeauthoryear{Johnson}{Johnson}{2020}]%
        {johnson2020apparent}
\bibfield{author}{\bibinfo{person}{Khari Johnson}.}
  \bibinfo{year}{2020}\natexlab{}.
\newblock \showarticletitle{Apparent racial bias found in Twitter photo
  algorithm}.
\newblock \bibinfo{journal}{\emph{VentureBeat}} (\bibinfo{date}{Sep}
  \bibinfo{year}{2020}).
\newblock
\urldef\tempurl%
\url{https://venturebeat.com/2020/09/20/apparent-racial-bias-found-in-twitter-photo-algorithm/}
\showURL{%
\tempurl}


\bibitem[\protect\citeauthoryear{Kang}{Kang}{2013a}]%
        {Kang2013}
\bibfield{author}{\bibinfo{person}{Inkoo Kang}.}
  \bibinfo{year}{2013}\natexlab{a}.
\newblock \showarticletitle{Businesses: ‘Yelp is the thug of the
  Internet’}.
\newblock \bibinfo{journal}{\emph{\textit{MuckRock}}} (\bibinfo{year}{2013}).
\newblock
\urldef\tempurl%
\url{https://www.muckrock.com/news/archives/2013/jan/23/businesses-yelp-thug-of-the-internet.}
\showURL{%
\tempurl}


\bibitem[\protect\citeauthoryear{Kang}{Kang}{2013b}]%
        {kang2013yelp}
\bibfield{author}{\bibinfo{person}{Inkoo Kang}.}
  \bibinfo{year}{2013}\natexlab{b}.
\newblock \showarticletitle{Read nearly 700 FTC complaints regarding Yelp}.
\newblock \bibinfo{journal}{\emph{MuckRock}} (\bibinfo{date}{Jan}
  \bibinfo{year}{2013}).
\newblock
\urldef\tempurl%
\url{https://www.muckrock.com/news/archives/2013/jan/23/businesses-yelp-thug-of-the-internet/}
\showURL{%
\tempurl}


\bibitem[\protect\citeauthoryear{Karizat, Delmonaco, Eslami, and
  Andalibi}{Karizat et~al\mbox{.}}{2021}]%
        {KARIZAT2021TikTok}
\bibfield{author}{\bibinfo{person}{Nadia Karizat}, \bibinfo{person}{Daniel
  Delmonaco}, \bibinfo{person}{Motahhare Eslami}, {and}
  \bibinfo{person}{Nazanin Andalibi}.} \bibinfo{year}{2021}\natexlab{}.
\newblock \showarticletitle{Algorithmic folk theories and identity: How TikTok
  users co-produce Knowledge of identity and engage in algorithmic resistance}.
\newblock \bibinfo{journal}{\emph{Proceedings of the ACM on Human-Computer
  Interaction}} \bibinfo{number}{CSCW2} (\bibinfo{year}{2021}),
  \bibinfo{pages}{forthcoming}.
\newblock


\bibitem[\protect\citeauthoryear{Kulshrestha, Eslami, Messias, Zafar, Ghosh,
  Gummadi, and Karahalios}{Kulshrestha et~al\mbox{.}}{2017}]%
        {kulshrestha2017quantifying}
\bibfield{author}{\bibinfo{person}{Juhi Kulshrestha},
  \bibinfo{person}{Motahhare Eslami}, \bibinfo{person}{Johnnatan Messias},
  \bibinfo{person}{Muhammad~Bilal Zafar}, \bibinfo{person}{Saptarshi Ghosh},
  \bibinfo{person}{Krishna~P Gummadi}, {and} \bibinfo{person}{Karrie
  Karahalios}.} \bibinfo{year}{2017}\natexlab{}.
\newblock \showarticletitle{Quantifying search bias: Investigating sources of
  bias for political searches in social media}. In
  \bibinfo{booktitle}{\emph{Proceedings of the 2017 ACM Conference on Computer
  Supported Cooperative Work and Social Computing}}. \bibinfo{pages}{417--432}.
\newblock


\bibitem[\protect\citeauthoryear{Lakkaraju, Kamar, Caruana, and
  Horvitz}{Lakkaraju et~al\mbox{.}}{2017}]%
        {lakkaraju2017identifying}
\bibfield{author}{\bibinfo{person}{Himabindu Lakkaraju}, \bibinfo{person}{Ece
  Kamar}, \bibinfo{person}{Rich Caruana}, {and} \bibinfo{person}{Eric
  Horvitz}.} \bibinfo{year}{2017}\natexlab{}.
\newblock \showarticletitle{Identifying unknown unknowns in the open world:
  Representations and policies for guided exploration}. In
  \bibinfo{booktitle}{\emph{Proceedings of the AAAI Conference on Artificial
  Intelligence}}, Vol.~\bibinfo{volume}{31}.
\newblock


\bibitem[\protect\citeauthoryear{Lee and Singh}{Lee and Singh}{2021}]%
        {lee2020landscape}
\bibfield{author}{\bibinfo{person}{Michelle Seng~Ah Lee} {and}
  \bibinfo{person}{Jatinder Singh}.} \bibinfo{year}{2021}\natexlab{}.
\newblock \showarticletitle{The landscape and gaps in open source fairness
  toolkits}. In \bibinfo{booktitle}{\emph{Proceedings of the 2021 CHI
  Conference on Human Factors in Computing Systems}}. \bibinfo{pages}{1--13}.
\newblock


\bibitem[\protect\citeauthoryear{Leskovec, Backstrom, and Kleinberg}{Leskovec
  et~al\mbox{.}}{2009}]%
        {leskovec2009meme}
\bibfield{author}{\bibinfo{person}{Jure Leskovec}, \bibinfo{person}{Lars
  Backstrom}, {and} \bibinfo{person}{Jon Kleinberg}.}
  \bibinfo{year}{2009}\natexlab{}.
\newblock \showarticletitle{Meme-tracking and the dynamics of the news cycle}.
  In \bibinfo{booktitle}{\emph{Proceedings of the 15th ACM SIGKDD International
  Conference on Knowledge Discovery and Data Mining}}.
  \bibinfo{pages}{497–506}.
\newblock
\urldef\tempurl%
\url{https://doi.org/10.1145/1557019.1557077}
\showDOI{\tempurl}


\bibitem[\protect\citeauthoryear{Liu, Guerra, Fung, Matute, Kamar, and
  Lasecki}{Liu et~al\mbox{.}}{2020}]%
        {liu2020towards}
\bibfield{author}{\bibinfo{person}{Anthony Liu}, \bibinfo{person}{Santiago
  Guerra}, \bibinfo{person}{Isaac Fung}, \bibinfo{person}{Gabriel Matute},
  \bibinfo{person}{Ece Kamar}, {and} \bibinfo{person}{Walter Lasecki}.}
  \bibinfo{year}{2020}\natexlab{}.
\newblock \showarticletitle{Towards hybrid human-AI workflows for unknown
  unknown detection}. In \bibinfo{booktitle}{\emph{Proceedings of The Web
  Conference 2020}}. \bibinfo{pages}{2432--2442}.
\newblock


\bibitem[\protect\citeauthoryear{Madaio, Stark, Wortman~Vaughan, and
  Wallach}{Madaio et~al\mbox{.}}{2020}]%
        {madaio2020co}
\bibfield{author}{\bibinfo{person}{Michael~A Madaio}, \bibinfo{person}{Luke
  Stark}, \bibinfo{person}{Jennifer Wortman~Vaughan}, {and}
  \bibinfo{person}{Hanna Wallach}.} \bibinfo{year}{2020}\natexlab{}.
\newblock \showarticletitle{Co-designing checklists to understand
  organizational challenges and opportunities around fairness in AI}. In
  \bibinfo{booktitle}{\emph{Proceedings of the 2020 CHI Conference on Human
  Factors in Computing Systems}}. \bibinfo{pages}{1--14}.
\newblock


\bibitem[\protect\citeauthoryear{Mehta}{Mehta}{2021}]%
        {mehta2021why}
\bibfield{author}{\bibinfo{person}{Ivan Mehta}.}
  \bibinfo{year}{2021}\natexlab{}.
\newblock \showarticletitle{Why Twitter's image cropping algorithm appears to
  have white bias}.
\newblock \bibinfo{journal}{\emph{TNW | Neural}} (\bibinfo{date}{Mar}
  \bibinfo{year}{2021}).
\newblock
\urldef\tempurl%
\url{https://thenextweb.com/news/why-twitters-image-cropping-algorithm-appears-to-have-white-bias}
\showURL{%
\tempurl}


\bibitem[\protect\citeauthoryear{Metz}{Metz}{2019}]%
        {metz2019nerd}
\bibfield{author}{\bibinfo{person}{Cade Metz}.}
  \bibinfo{year}{2019}\natexlab{}.
\newblock \showarticletitle{'Nerd,' 'Nonsmoker,' 'Wrongdoer': How Might A.I.
  Label You?}
\newblock \bibinfo{journal}{\emph{The New York Times}} (\bibinfo{date}{Sep}
  \bibinfo{year}{2019}).
\newblock
\urldef\tempurl%
\url{https://www.nytimes.com/2019/09/20/arts/design/imagenet-trevor-paglen-ai-facial-recognition.html}
\showURL{%
\tempurl}


\bibitem[\protect\citeauthoryear{Mincy}{Mincy}{1993}]%
        {mincy1993urban}
\bibfield{author}{\bibinfo{person}{Ronald~B Mincy}.}
  \bibinfo{year}{1993}\natexlab{}.
\newblock \showarticletitle{The Urban Institute audit studies: Their research
  and policy context}.
\newblock In \bibinfo{booktitle}{\emph{Clear and convincing evidence:
  Measurement of discrimination in America}},
  \bibfield{editor}{\bibinfo{person}{M.~Fix} {and} \bibinfo{person}{R.~J.
  Struyk}} (Eds.). \bibinfo{publisher}{Urban Institute Press},
  \bibinfo{pages}{165--86}.
\newblock


\bibitem[\protect\citeauthoryear{Ni, Yang, Lin, and He}{Ni
  et~al\mbox{.}}{2017}]%
        {ni2017reducing}
\bibfield{author}{\bibinfo{person}{Mengjun Ni}, \bibinfo{person}{Jing Yang},
  \bibinfo{person}{Xin Lin}, {and} \bibinfo{person}{Liang He}.}
  \bibinfo{year}{2017}\natexlab{}.
\newblock \showarticletitle{Reducing unknown unknowns with guidance in image
  caption}. In \bibinfo{booktitle}{\emph{International Conference on Artificial
  Neural Networks}}. Springer, \bibinfo{pages}{547--555}.
\newblock


\bibitem[\protect\citeauthoryear{Noble}{Noble}{2018}]%
        {noble2018algorithms}
\bibfield{author}{\bibinfo{person}{Safiya~Umoja Noble}.}
  \bibinfo{year}{2018}\natexlab{}.
\newblock \bibinfo{booktitle}{\emph{Algorithms of oppression: How search
  engines reinforce racism}}.
\newblock \bibinfo{publisher}{NYU Press}.
\newblock


\bibitem[\protect\citeauthoryear{Olson}{Olson}{2018}]%
        {olson2018algorithm}
\bibfield{author}{\bibinfo{person}{Parmy Olson}.}
  \bibinfo{year}{2018}\natexlab{}.
\newblock \showarticletitle{The algorithm that helped google translate become
  sexist}.
\newblock \bibinfo{journal}{\emph{Forbes}} (\bibinfo{date}{Feb.}
  \bibinfo{year}{2018}).
\newblock
\urldef\tempurl%
\url{https://www.forbes.com/sites/parmyolson/2018/02/15/the-algorithm-that-helped-google-translate-become-sexist/?sh=6c1d0807daa2}
\showURL{%
\tempurl}


\bibitem[\protect\citeauthoryear{Ozment}{Ozment}{2004}]%
        {ozment2004bug}
\bibfield{author}{\bibinfo{person}{Andy Ozment}.}
  \bibinfo{year}{2004}\natexlab{}.
\newblock \showarticletitle{Bug auctions: Vulnerability markets reconsidered}.
  In \bibinfo{booktitle}{\emph{Third Workshop on the Economics of Information
  Security}}. \bibinfo{pages}{19--26}.
\newblock


\bibitem[\protect\citeauthoryear{Palmiotto}{Palmiotto}{2020}]%
        {palmiotto2020yelp}
\bibfield{author}{\bibinfo{person}{Francesca Palmiotto}.}
  \bibinfo{year}{2020}\natexlab{}.
\newblock \bibinfo{title}{Challenging automated filtering systems - The case of
  Yelp}.
\newblock
\newblock
\urldef\tempurl%
\url{https://ai-laws.org/en/2020/05/challenging-automated-filtering-systems-the-case-of-yelp/}
\showURL{%
\tempurl}


\bibitem[\protect\citeauthoryear{Ragin and Becker}{Ragin and Becker}{1992}]%
        {ragin1992case}
\bibfield{author}{\bibinfo{person}{Charles~C. Ragin} {and}
  \bibinfo{person}{Howard~S. Becker}.} \bibinfo{year}{1992}\natexlab{}.
\newblock \bibinfo{booktitle}{\emph{What is a case?: Exploring the foundations
  of social inquiry}}.
\newblock \bibinfo{publisher}{Cambridge University Press}.
\newblock


\bibitem[\protect\citeauthoryear{Raji and Buolamwini}{Raji and
  Buolamwini}{2019}]%
        {raji2019actionable}
\bibfield{author}{\bibinfo{person}{Inioluwa~Deborah Raji} {and}
  \bibinfo{person}{Joy Buolamwini}.} \bibinfo{year}{2019}\natexlab{}.
\newblock \showarticletitle{Actionable auditing: Investigating the impact of
  publicly naming biased performance results of commercial AI products}. In
  \bibinfo{booktitle}{\emph{Proceedings of the 2019 AAAI/ACM Conference on AI,
  Ethics, and Society}}. \bibinfo{pages}{429--435}.
\newblock


\bibitem[\protect\citeauthoryear{Raji, Smart, White, Mitchell, Gebru,
  Hutchinson, Smith-Loud, Theron, and Barnes}{Raji et~al\mbox{.}}{2020}]%
        {raji2020closing}
\bibfield{author}{\bibinfo{person}{Inioluwa~Deborah Raji},
  \bibinfo{person}{Andrew Smart}, \bibinfo{person}{Rebecca~N White},
  \bibinfo{person}{Margaret Mitchell}, \bibinfo{person}{Timnit Gebru},
  \bibinfo{person}{Ben Hutchinson}, \bibinfo{person}{Jamila Smith-Loud},
  \bibinfo{person}{Daniel Theron}, {and} \bibinfo{person}{Parker Barnes}.}
  \bibinfo{year}{2020}\natexlab{}.
\newblock \showarticletitle{Closing the AI accountability gap: Defining an
  end-to-end framework for internal algorithmic auditing}. In
  \bibinfo{booktitle}{\emph{Proceedings of the 2020 Conference on Fairness,
  Accountability, and Transparency}}. \bibinfo{pages}{33--44}.
\newblock


\bibitem[\protect\citeauthoryear{Robertson, Jiang, Joseph, Friedland, Lazer,
  and Wilson}{Robertson et~al\mbox{.}}{2018}]%
        {robertson2018auditing}
\bibfield{author}{\bibinfo{person}{Ronald~E Robertson}, \bibinfo{person}{Shan
  Jiang}, \bibinfo{person}{Kenneth Joseph}, \bibinfo{person}{Lisa Friedland},
  \bibinfo{person}{David Lazer}, {and} \bibinfo{person}{Christo Wilson}.}
  \bibinfo{year}{2018}\natexlab{}.
\newblock \showarticletitle{Auditing partisan audience bias within google
  search}.
\newblock \bibinfo{journal}{\emph{Proceedings of the ACM on Human-Computer
  Interaction}} \bibinfo{volume}{2}, \bibinfo{number}{CSCW}
  (\bibinfo{year}{2018}), \bibinfo{pages}{1--22}.
\newblock


\bibitem[\protect\citeauthoryear{Romano}{Romano}{2019}]%
        {romano2019group}
\bibfield{author}{\bibinfo{person}{Aja Romano}.}
  \bibinfo{year}{2019}\natexlab{}.
\newblock \showarticletitle{A group of YouTubers is trying to prove the site
  systematically demonetizes Queer content}.
\newblock \bibinfo{journal}{\emph{Vox}} (\bibinfo{date}{Oct.}
  \bibinfo{year}{2019}).
\newblock
\urldef\tempurl%
\url{https://www. vox.
  com/culture/2019/10/10/20893258/youtube-lgbtq-censorship-demonetization-nerd-city-algorithm-report}
\showURL{%
\tempurl}


\bibitem[\protect\citeauthoryear{Sandvig, Hamilton, Karahalios, and
  Langbort}{Sandvig et~al\mbox{.}}{2014}]%
        {sandvig2014auditing}
\bibfield{author}{\bibinfo{person}{Christian Sandvig}, \bibinfo{person}{Kevin
  Hamilton}, \bibinfo{person}{Karrie Karahalios}, {and} \bibinfo{person}{Cedric
  Langbort}.} \bibinfo{year}{2014}\natexlab{}.
\newblock \showarticletitle{Auditing algorithms: Research methods for detecting
  discrimination on internet platforms}.
\newblock \bibinfo{journal}{\emph{Data and Discrimination: Converting Critical
  Concerns into Productive Inquiry}} (\bibinfo{year}{2014}).
\newblock


\bibitem[\protect\citeauthoryear{Sanjay}{Sanjay}{2020}]%
        {sanjay2020twitter}
\bibfield{author}{\bibinfo{person}{Satviki Sanjay}.}
  \bibinfo{year}{2020}\natexlab{}.
\newblock \showarticletitle{Twitter’s AI Revealed to Have Racial Bias After a
  Viral Photo Experiment}.
\newblock \bibinfo{journal}{\emph{Vice}} (\bibinfo{date}{Sep}
  \bibinfo{year}{2020}).
\newblock
\urldef\tempurl%
\url{https://www.vice.com/en/article/93547v/twitter-ai-algorithm-has-racial-bias}
\showURL{%
\tempurl}


\bibitem[\protect\citeauthoryear{Schechter}{Schechter}{2002}]%
        {schechter2002buy}
\bibfield{author}{\bibinfo{person}{Stuart Schechter}.}
  \bibinfo{year}{2002}\natexlab{}.
\newblock \showarticletitle{How to buy better testing using competition to get
  the most security and robustness for your dollar}. In
  \bibinfo{booktitle}{\emph{International Conference on Infrastructure
  Security}}. Springer, \bibinfo{pages}{73--87}.
\newblock


\bibitem[\protect\citeauthoryear{Scott}{Scott}{1985}]%
        {scott1985weapons}
\bibfield{author}{\bibinfo{person}{James~C Scott}.}
  \bibinfo{year}{1985}\natexlab{}.
\newblock \bibinfo{booktitle}{\emph{Weapons of the weak: Everyday forms of
  peasant resistance}}.
\newblock \bibinfo{publisher}{Yale University Press}.
\newblock


\bibitem[\protect\citeauthoryear{Seaver}{Seaver}{2017}]%
        {seaver2017algorithms}
\bibfield{author}{\bibinfo{person}{Nick Seaver}.}
  \bibinfo{year}{2017}\natexlab{}.
\newblock \showarticletitle{Algorithms as culture: Some tactics for the
  ethnography of algorithmic systems}.
\newblock \bibinfo{journal}{\emph{Big Data \& Society}} \bibinfo{volume}{4},
  \bibinfo{number}{2} (\bibinfo{year}{2017}).
\newblock
\urldef\tempurl%
\url{https://doi.org/doi:10.1177/2053951717738104}
\showDOI{\tempurl}


\bibitem[\protect\citeauthoryear{Seaver}{Seaver}{2019}]%
        {seaver2019knowing}
\bibfield{author}{\bibinfo{person}{Nick Seaver}.}
  \bibinfo{year}{2019}\natexlab{}.
\newblock \showarticletitle{Knowing algorithms}.
\newblock \bibinfo{journal}{\emph{Digital STS}} (\bibinfo{year}{2019}),
  \bibinfo{pages}{412--422}.
\newblock


\bibitem[\protect\citeauthoryear{Selbst, Boyd, Friedler, Venkatasubramanian,
  and Vertesi}{Selbst et~al\mbox{.}}{2019}]%
        {selbst2019fairness}
\bibfield{author}{\bibinfo{person}{Andrew~D Selbst}, \bibinfo{person}{Danah
  Boyd}, \bibinfo{person}{Sorelle~A Friedler}, \bibinfo{person}{Suresh
  Venkatasubramanian}, {and} \bibinfo{person}{Janet Vertesi}.}
  \bibinfo{year}{2019}\natexlab{}.
\newblock \showarticletitle{Fairness and abstraction in sociotechnical
  systems}. In \bibinfo{booktitle}{\emph{Proceedings of the Conference on
  Fairness, Accountability, and Transparency}}. \bibinfo{pages}{59--68}.
\newblock


\bibitem[\protect\citeauthoryear{Shifman}{Shifman}{2013}]%
        {shifman2013memes}
\bibfield{author}{\bibinfo{person}{Limor Shifman}.}
  \bibinfo{year}{2013}\natexlab{}.
\newblock \showarticletitle{Memes in a digital world: Reconciling with a
  conceptual troublemaker}.
\newblock \bibinfo{journal}{\emph{Journal of Computer-Mediated Communication}}
  \bibinfo{volume}{18}, \bibinfo{number}{3} (\bibinfo{year}{2013}),
  \bibinfo{pages}{362--377}.
\newblock


\bibitem[\protect\citeauthoryear{Simpson and Semaan}{Simpson and
  Semaan}{2021}]%
        {simpson2021you}
\bibfield{author}{\bibinfo{person}{Ellen Simpson} {and} \bibinfo{person}{Bryan
  Semaan}.} \bibinfo{year}{2021}\natexlab{}.
\newblock \showarticletitle{For you, or For ``you''? Everyday LGBTQ+ encounters
  with TikTok}.
\newblock \bibinfo{journal}{\emph{Proceedings of the ACM on Human-Computer
  Interaction}} \bibinfo{volume}{4}, \bibinfo{number}{CSCW3}
  (\bibinfo{year}{2021}), \bibinfo{pages}{1--34}.
\newblock


\bibitem[\protect\citeauthoryear{Suchman}{Suchman}{1987}]%
        {suchman1987plans}
\bibfield{author}{\bibinfo{person}{Lucy~A Suchman}.}
  \bibinfo{year}{1987}\natexlab{}.
\newblock \bibinfo{booktitle}{\emph{Plans and situated actions: The problem of
  human-machine communication}}.
\newblock \bibinfo{publisher}{Cambridge University Press}.
\newblock


\bibitem[\protect\citeauthoryear{Sweeney}{Sweeney}{2013}]%
        {sweeney2013discrimination}
\bibfield{author}{\bibinfo{person}{Latanya Sweeney}.}
  \bibinfo{year}{2013}\natexlab{}.
\newblock \showarticletitle{Discrimination in online ad delivery}.
\newblock \bibinfo{journal}{\emph{Queue}} \bibinfo{volume}{11},
  \bibinfo{number}{3} (\bibinfo{year}{2013}), \bibinfo{pages}{10--29}.
\newblock


\bibitem[\protect\citeauthoryear{Theis and Wang}{Theis and Wang}{2018}]%
        {theis2018neural}
\bibfield{author}{\bibinfo{person}{Lucas Theis} {and} \bibinfo{person}{Zehan
  Wang}.} \bibinfo{year}{2018}\natexlab{}.
\newblock \bibinfo{title}{Speedy Neural Networks for Smart Auto-Cropping of
  Images}.
\newblock
\newblock
\urldef\tempurl%
\url{https://blog.twitter.com/engineering/en_us/topics/infrastructure/2018/Smart-Auto-Cropping-of-Images.html}
\showURL{%
\tempurl}


\bibitem[\protect\citeauthoryear{Tucker}{Tucker}{2016}]%
        {tucker2016uber}
\bibfield{author}{\bibinfo{person}{Harry Tucker}.}
  \bibinfo{year}{2016}\natexlab{}.
\newblock \showarticletitle{Australian Uber drivers say the company is
  manipulating their ratings to boost its fees}.
\newblock \bibinfo{journal}{\emph{Businessinsider}} (\bibinfo{date}{May}
  \bibinfo{year}{2016}).
\newblock
\urldef\tempurl%
\url{https://www.businessinsider.com.au/australian-uber-drivers-say-the-company-is-manipulating-their-ratings-to-boost-the-companys-fees-2016-5}
\showURL{%
\tempurl}


\bibitem[\protect\citeauthoryear{Vandenhof and Law}{Vandenhof and Law}{2019}]%
        {vandenhof2019contradict}
\bibfield{author}{\bibinfo{person}{Colin Vandenhof} {and}
  \bibinfo{person}{Edith Law}.} \bibinfo{year}{2019}\natexlab{}.
\newblock \showarticletitle{Contradict the Machine: A Hybrid Approach to
  Identifying Unknown Unknowns}. In \bibinfo{booktitle}{\emph{Proceedings of
  the 18th International Conference on Autonomous Agents and MultiAgent
  Systems}}. \bibinfo{pages}{2238--2240}.
\newblock


\bibitem[\protect\citeauthoryear{Veale, Van~Kleek, and Binns}{Veale
  et~al\mbox{.}}{2018}]%
        {veale2018fairness}
\bibfield{author}{\bibinfo{person}{Michael Veale}, \bibinfo{person}{Max
  Van~Kleek}, {and} \bibinfo{person}{Reuben Binns}.}
  \bibinfo{year}{2018}\natexlab{}.
\newblock \showarticletitle{Fairness and accountability design needs for
  algorithmic support in high-stakes public sector decision-making}. In
  \bibinfo{booktitle}{\emph{Proceedings of the 2018 CHI Conference on Human
  Factors in Computing Systems}}. \bibinfo{pages}{1--14}.
\newblock


\bibitem[\protect\citeauthoryear{Velkova and Kaun}{Velkova and Kaun}{2019}]%
        {velkova2019algorithmic}
\bibfield{author}{\bibinfo{person}{Julia Velkova} {and} \bibinfo{person}{Anne
  Kaun}.} \bibinfo{year}{2019}\natexlab{}.
\newblock \showarticletitle{Algorithmic resistance: Media practices and the
  politics of repair}.
\newblock \bibinfo{journal}{\emph{Information, Communication \& Society}}
  (\bibinfo{year}{2019}), \bibinfo{pages}{1--18}.
\newblock


\bibitem[\protect\citeauthoryear{Vigdor}{Vigdor}{2019}]%
        {vigdor2019apple}
\bibfield{author}{\bibinfo{person}{Neil Vigdor}.}
  \bibinfo{year}{2019}\natexlab{}.
\newblock \showarticletitle{Apple card investigated after gender discrimination
  complaints}.
\newblock \bibinfo{journal}{\emph{The New York Times}} (\bibinfo{date}{Nov.}
  \bibinfo{year}{2019}).
\newblock
\urldef\tempurl%
\url{https://www.nytimes.com/2019/11/10/business/Apple-credit-card-investigation.html}
\showURL{%
\tempurl}


\bibitem[\protect\citeauthoryear{Vincent}{Vincent}{2016}]%
        {vincent2016chatbot}
\bibfield{author}{\bibinfo{person}{James Vincent}.}
  \bibinfo{year}{2016}\natexlab{}.
\newblock \showarticletitle{Twitter taught Microsoft's AI chatbot to be a
  racist asshole in less than a day}.
\newblock \bibinfo{journal}{\emph{The Verge}} (\bibinfo{date}{Mar}
  \bibinfo{year}{2016}).
\newblock
\urldef\tempurl%
\url{https://www.theverge.com/2016/3/24/11297050/tay-microsoft-chatbot-racist}
\showURL{%
\tempurl}


\bibitem[\protect\citeauthoryear{Vincent}{Vincent}{2018}]%
        {vincent2018google}
\bibfield{author}{\bibinfo{person}{James Vincent}.}
  \bibinfo{year}{2018}\natexlab{}.
\newblock \showarticletitle{Google ‘fixed’ its racist algorithm by removing
  gorillas from its image-labeling tech}.
\newblock \bibinfo{journal}{\emph{The Verge}} (\bibinfo{date}{Jan.}
  \bibinfo{year}{2018}).
\newblock
\urldef\tempurl%
\url{https://www.theverge.com/2018/1/12/16882408/google-racist-gorillas-photo-recognition-algorithm-ai}
\showURL{%
\tempurl}


\bibitem[\protect\citeauthoryear{Vincent, Li, Tilly, Chancellor, and
  Hecht}{Vincent et~al\mbox{.}}{2021}]%
        {vincent2021data}
\bibfield{author}{\bibinfo{person}{Nicholas Vincent}, \bibinfo{person}{Hanlin
  Li}, \bibinfo{person}{Nicole Tilly}, \bibinfo{person}{Stevie Chancellor},
  {and} \bibinfo{person}{Brent Hecht}.} \bibinfo{year}{2021}\natexlab{}.
\newblock \showarticletitle{Data leverage: A framework for empowering the
  public in its relationship with technology companies}. In
  \bibinfo{booktitle}{\emph{Proceedings of the 2021 ACM Conference on Fairness,
  Accountability, and Transparency}}. \bibinfo{pages}{215--227}.
\newblock


\bibitem[\protect\citeauthoryear{Wang}{Wang}{2020}]%
        {wang2020calculating}
\bibfield{author}{\bibinfo{person}{Shuaishuai Wang}.}
  \bibinfo{year}{2020}\natexlab{}.
\newblock \showarticletitle{Calculating dating goals: data gaming and
  algorithmic sociality on Blued, a Chinese gay dating app}.
\newblock \bibinfo{journal}{\emph{Information, Communication \& Society}}
  \bibinfo{volume}{23}, \bibinfo{number}{2} (\bibinfo{year}{2020}),
  \bibinfo{pages}{181--197}.
\newblock


\bibitem[\protect\citeauthoryear{Weick}{Weick}{1995}]%
        {weick1995sensemaking}
\bibfield{author}{\bibinfo{person}{Karl~E Weick}.}
  \bibinfo{year}{1995}\natexlab{}.
\newblock \bibinfo{booktitle}{\emph{Sensemaking in organizations}}.
  Vol.~\bibinfo{volume}{3}.
\newblock \bibinfo{publisher}{Sage}.
\newblock


\bibitem[\protect\citeauthoryear{Wiggers}{Wiggers}{2020}]%
        {wiggers2020google}
\bibfield{author}{\bibinfo{person}{Kyle Wiggers}.}
  \bibinfo{year}{2020}\natexlab{}.
\newblock \showarticletitle{Google debuts AI in Google Translate that addresses
  gender bias}.
\newblock \bibinfo{journal}{\emph{VentureBeat}} (\bibinfo{date}{Apr}
  \bibinfo{year}{2020}).
\newblock
\urldef\tempurl%
\url{https://venturebeat.com/2020/04/22/google-debuts-ai-in-google-translate-that-addresses-gender-bias}
\showURL{%
\tempurl}


\bibitem[\protect\citeauthoryear{Xiao, Metaxa, Park, Karahalios, and
  Salehi}{Xiao et~al\mbox{.}}{2020}]%
        {xiao2020random}
\bibfield{author}{\bibinfo{person}{Sijia Xiao}, \bibinfo{person}{Dana{\"e}
  Metaxa}, \bibinfo{person}{Joon~Sung Park}, \bibinfo{person}{Karrie
  Karahalios}, {and} \bibinfo{person}{Niloufar Salehi}.}
  \bibinfo{year}{2020}\natexlab{}.
\newblock \showarticletitle{Random, messy, funny, raw: Finstas as intimate
  reconfigurations of social media}. In \bibinfo{booktitle}{\emph{Proceedings
  of the 2020 CHI Conference on Human Factors in Computing Systems}}.
  \bibinfo{pages}{1--13}.
\newblock


\bibitem[\protect\citeauthoryear{Yelp}{Yelp}{2010}]%
        {yelp}
\bibfield{author}{\bibinfo{person}{Yelp}.} \bibinfo{year}{2010}\natexlab{}.
\newblock \bibinfo{title}{Yelp’s Review Filter Explained}.
\newblock
\newblock
\urldef\tempurl%
\url{https://blog.yelp.com/2010/03/yelp-review-filter-explained}
\showURL{%
\tempurl}


\bibitem[\protect\citeauthoryear{Young, Magassa, and Friedman}{Young
  et~al\mbox{.}}{2019}]%
        {young2019toward}
\bibfield{author}{\bibinfo{person}{Meg Young}, \bibinfo{person}{Lassana
  Magassa}, {and} \bibinfo{person}{Batya Friedman}.}
  \bibinfo{year}{2019}\natexlab{}.
\newblock \showarticletitle{Toward inclusive tech policy design: A method for
  underrepresented voices to strengthen tech policy documents}.
\newblock \bibinfo{journal}{\emph{Ethics and Information Technology}}
  \bibinfo{volume}{21}, \bibinfo{number}{2} (\bibinfo{year}{2019}),
  \bibinfo{pages}{89--103}.
\newblock


\bibitem[\protect\citeauthoryear{Zou and Schiebinger}{Zou and
  Schiebinger}{2018}]%
        {zou2018ai}
\bibfield{author}{\bibinfo{person}{James Zou} {and} \bibinfo{person}{Londa
  Schiebinger}.} \bibinfo{year}{2018}\natexlab{}.
\newblock \showarticletitle{AI can be sexist and racist—it’s time to make
  it fair}.
\newblock \bibinfo{journal}{\emph{Nature}}  \bibinfo{volume}{559}
  (\bibinfo{year}{2018}), \bibinfo{pages}{324--326}.
\newblock


\bibitem[\protect\citeauthoryear{Ángel Alexander~Cabrera, Druck, Hong, and
  Perer}{Ángel Alexander~Cabrera et~al\mbox{.}}{2021}]%
        {Cabrera2021Deblinder}
\bibfield{author}{\bibinfo{person}{Ángel Alexander~Cabrera},
  \bibinfo{person}{Abraham Druck}, \bibinfo{person}{Jason~I Hong}, {and}
  \bibinfo{person}{Adam Perer}.} \bibinfo{year}{2021}\natexlab{}.
\newblock \showarticletitle{Discovering and Validating AI Errors With
  Crowdsourced Failure Reports}.
\newblock \bibinfo{journal}{\emph{Proceedings of the ACM Conference on Computer
  Supported Cooperative Work, CSCW}}  \bibinfo{volume}{5}
  (\bibinfo{year}{2021}), \bibinfo{pages}{forthcoming}.
\newblock
\urldef\tempurl%
\url{https://doi.org/10.1145/3479569}
\showDOI{\tempurl}


\end{thebibliography}


\end{document}